\documentclass{birkjour}
\usepackage{orcidlink}
\usepackage{xcolor}
\definecolor{accentcolor}{RGB}{0, 51, 102}
\usepackage{url}
\usepackage{graphicx}
\usepackage{amsmath}
\usepackage{amsfonts}
\usepackage{physics}
\usepackage{hyperref} 
\usepackage{nicefrac}
\usepackage{doi}
\usepackage{slashed}
\usepackage{booktabs} 
\usepackage[english]{babel}
\usepackage[autostyle, english = american]{csquotes}
\MakeOuterQuote{"}
\usepackage{mathpazo}

\usepackage[style=numeric,sorting=none]{biblatex} 

\RequirePackage[all]{xy}

\theoremstyle{definition}

\theoremstyle{remark}

\numberwithin{equation}{section}

\newcommand{\BibTeX}{B\kern-0.1emi\kern-0.017emb\kern-0.15em\TeX}
\newcommand{\XYpic}{$\mathrm{X\kern-0.3em\raisebox{-0.18em}{Y}}$-$\mathrm{pic}\,$}

\newcommand{\cl}{C \kern -0.1em \ell}  

\newcommand{\BR}{\mathbb{R}}

\newcommand{\STA}{\mathbb{G}_{1,3}}

\newcommand{\APS}{\mathbb{G}_3}

\newcommand{\Spin}{\mathrm{Spin}}

\newcommand{\SU}{\mathrm{SU}}

\newcommand{\SL}{\mathrm{SL}}

\newcommand{\la}{\lambda}

\newcommand{\al}{\alpha}
\newcommand{\be}{\beta}

\newcommand{\Le}{\mathrm{L}}
\newcommand{\Ri}{\mathrm{R}}
\newcommand{\Fo}{\mathrm{F}}
\newcommand{\Ba}{\mathrm{B}}
\newcommand{\jj}{\mathrm{J}}
\newcommand{\kk}{\mathrm{K}}
\newcommand{\mm}{\mathrm{M}}
\newcommand{\nn}{\mathrm{N}}
\newcommand{\sip}{\!:\!}
\newcommand{\sop}{\mathbin{\hbox{$\wedge$}\kern-0.457em\raisebox{0ex}{$\cdot$}}}

\newcommand{\cproj}[1]{\left<#1\right>_{0\oplus3}}
\newcommand{\gproj}[2]{\left<#1\right>_{#2}}
\newcommand{\dagg}{\dagger}
\newcommand{\LP}[2]{\hat{B}\widetilde{\lambda}_{#1}\lambda_{#2}}
\newcommand{\mLP}[2]{\hat{B}^\dagger\lambda_{#1}^\dagger\lambda_{#2}^-}

\newcommand{\SAb}[1]{\left[{#1}\right]}
\newcommand{\upSAb}[3]{\left[{#1}\right]^{{#2} {#3}}}
\newcommand{\downSAb}[3]{\left[{#1}\right]_{{#3} {#2}}}
\usepackage[normalem]{ulem}

\newcommand{\harpoon}{\overset{\rightharpoonup}}

\newcommand{\ed}{\end{document}}
\begin{document}

%
%
%
%
%
%
%
%
%
\makeatletter
\def\@setauthors{%
  \begingroup
  \trivlist\Large
  \centering          
  \@topsep30\p@\relax
  \advance\@topsep by -\baselineskip
  \item\relax
  \andify\authors
  \authors
  \endtrivlist
  \endgroup
  \vskip20pt
}
\makeatother

\title[Wigner-Covariance and Fields]
 {The Scattering Algebra of Physical Space: Wigner-Covariance and Fields}
\author[Moab Croft]{Moab Croft \orcidlink{0000-0003-4082-0428}} 
%
\address{%
Department of Physics, Illinois State University\\
Normal\\
IL, 61790\\
USA}
\email{mecroft@ilstu.edu}

\author[Neil Christensen]{Neil Christensen\orcidlink{0000-0002-2733-0857}}
%
\address{%
Department of Physics, Illinois State University\\
Normal\\
IL, 61790\\
USA}
\email{nchris3@ilstu.edu}

%

%
%
\date{\today}
\dedicatory{Last Revised:\\ September 15, 2026}

\begin{abstract}
Following previous work, the \textit{Algebra of Physical Space} (APS) is used to explore Wigner-covariance and spin/helicity fields within the \textit{Constructive Standard Model} (CSM) of Particle Physics. The spinor formalism of the APS is used to derive explicit Wigner-covariance of Lorentz spinors, and equivalencies with the CSM are demonstrated via the \textit{Scattering Algebra} (SA). Constructive fields for spin-$1/2$ and spin-$1$ are given in the APS and the necessary maps for Wigner-covariance are proposed. The forms of these spin fields are equivalent to Pauli spinors, thereby serving as a new bridge between the CSM and the study of Quantum Information. It is further seen that spin fields of the APS are equivalent to the spin fields in the CSM. Sample Lagrangian densities of the CSM are analyzed using the methods herein, are given geometric interpretations, and are connected to traditional Pauli Theory. Finally, the $hf\overline{f}$ (Higgs and two massive fermions) Lagrangian density is used to determine the \textit{first} constructive scattering amplitude that is defined \textit{purely in terms of the APS}. Concerningly, this leads to an anti-Hermitian action term, and this surprise is later confirmed using traditional CSM techniques. Throughout this paper, the illuminating power of Geometric Algebra is clear: Everything has a geometric interpretation, and results can be accomplished matrix-free.
\end{abstract}
\label{Abstract}

\maketitle

\newpage

\section{Introduction}\label{Intro}

In a previous paper \cite{Croft2026}, the \textit{Constructive Standard Model} (CSM) of Particle Physics \cite{Christensen2018,Christensen2020,Christensen2024,2Christensen2024,3Christensen2024,4Christensen2024}---an in-development competitor to the traditional Feynman rules-based Standard Model, herein called the \textit{Feynman Standard Model}---was for the first time analyzed using the methods of \textit{Geometric Algebra}. The paper used the \textit{Algebra of Physical Space} (APS), $\APS$, and connected the traditional CSM algebra to the APS via what was called the \textit{Scattering Algebra} (SA). Through this connection new discoveries were made, such as: Realizing chiral structure as \textit{ray spinor structure}, realizing the CSM's spin spinors to be (chiral) \textit{Lorentz spinors} (Lorentz rotors scaled by $\sqrt{mc}$) thereby implying that no particle-spin information was contained, and the finding of a hitherto unused \textit{achiral} spin spinor due to the existence of a Hermitian Lorentz spinor. Moreover, the geometric insights of the APS gave newfound interpretations and allowed for a matrix-free and coordinate-free squaring of massive constructive amplitudes. 

Despite these significant advances, there remains much to be done. Lines of further research in the APS and SA include: Constructive amplitudes for all massless cases, amplitude-construction, the exploration of Wigner little group (WLG) methods, and constructive fields. This paper looks at the latter two. In SEC.~\ref{W}, WLG transformations for massive and massless Lorentz spinors within the APS---as well as their SA counterparts---will at last be discussed. In SEC.~\ref{F}, constructive fields will likewise be presented within the APS and then within the SA. As a sidenote, the algebraic form of these fields will serve as a first-bridge between the CSM and the discipline of Quantum Information. Also, despite the fields' existence in the APS, the SA equivalents of the massless spin-$1$ fields cannot yet be fully used as they interact with what is called the $x$-factor \cite{Christensen2023}, whose form within the SA has yet to be explicitly considered. In SEC.~\ref{T}, the results of all previous sections will be used to analyze and interpret the Lagrangian densities for two massive free terms, a two massive interacting term, and a massive-massless interacting term: $f\overline{f}$ (free massive fermion), $WW$ (free W-boson), $hf\overline{f}$ (Higgs and two massive fermions), $q\overline{q}W$ (W-boson and two quarks), and $l\nu W$ (W-boson, lepton, and neutrino). The $hf\overline{f}$ term will give the first constructive amplitude to be \textit{fully inside the APS}. The $hf\overline{f}$ term is a measure of geometric overlap with the timelike direction, while the $q\overline{q}W$ and $l\nu W$ are conceptually equivalent to the inner products between two paravectors (spacetime vectors). It will also be seen that the constructive amplitude for $hf\overline{f}$ leads to an anti-Hermitian action term. A verification using traditional CSM techniques confirms this. This paper concludes with a discussion of all covered results.

As this work directly follows \cite{Croft2026}, a read-through of the preceding paper is recommended. Nevertheless, all strictly necessary background will be covered by the appendices. Also in the appendices will be a proof of anti-Hermiticity relevant to SEC.~\ref{T} and a table of concurrence summarizing equivalencies between the CSM, the SA, and the APS.

\section{Wigner Little Groups and Lorentz Spinors}\label{W}

The discussion of \textit{Wigner little groups} (WLG), for either massive or massless particles, involves studying the properties of Lorentz spinors under Lorentz transformations. There are \textit{chiral} and \textit{achiral} Lorentz spinors. Both have massive and massless cases. This section will discuss the WLG for each subcase. At the section's end, the results will be injected into the Scattering Algebra (SA)---originally presented in \cite{Croft2026}---thereby showing equivalency with the results of the Constructive Standard Model (CSM) \cite{2Christensen2024}.

It will be seen that Lorentz spinors, in bijective equivalency with the spinors of the CSM, are \textit{Wigner-covariant}: An external Lorentz transformation of a Lorentz spinor defined at momentum $p$ is \textit{expressly equal} to an internal WLG transformation of a Lorentz spinor defined at the new momentum $\overline{p}$. For the reader's convenience, the Algebra of Physical Space equations showing Wigner-covariance are EQ.~\ref{EQW: Chiral Case}, EQ.~\ref{EQW: Achiral Case}, EQ.~\ref{EQW: Massless chiral case}, and EQ.~\ref{EQW: Massless achiral case}. The SA equations showing Wigner-covariance are EQ.~\ref{EQW: Massive SA Chiral Case}, EQ.~\ref{EQW: Massless SA Chiral Case}, EQ.~\ref{EQW: Massive SA achiral Case}, and EQ.~\ref{EQW: Massless SA achiral Case}.

\subsection{Massive Lorentz Spinors}

A massive \textit{chiral}\footnote{The chirality of Lorentz spinors is presented within the appendix of \cite{Croft2026}.} Lorentz spinor is defined as
\begin{equation}\label{EQW: Chiral Lorentz spinor}
    \la = \sqrt{mc}R_pL_a,
\end{equation}
where $m$ is the particle's mass, $L_a=\cosh{\eta/2}+\hat{a}\sinh{\eta/2}$ is a Lorentz boost with rapidity $\eta$ in the direction of the \textit{reference axis} $\hat{a}\in\APS^1$, and $R_p=\cos{\theta/2}+\hat{p}\wedge\hat{a}\sin{\theta/2}$ is the rotor that takes the reference axis $\hat{a}$ to the \textit{spatial momentum direction} $\hat{p}\in\APS^1$. A massive \textit{achiral} Lorentz spinor is defined as
\begin{equation}\label{EQW: Achiral Lorentz spinor}
    \la_p = \lambda R_p^\dagg = \sqrt{mc}L_p,
\end{equation}
where $L_p=\cosh{\eta/2}+\hat{p}\sinh{\eta/2}$ is a boost in the direction of the spatial momentum direction $\hat{p}$. Both chiral and achiral Lorentz spinors satisfy the definition of \textit{spacetime momentum},
\begin{equation}\label{EQW: Spacetime Momentum}
    p = \frac{1}{c}E+\mathbf{p}=\la\la^\dagg=\la_p\la_p^\dagg,
\end{equation}
for energy $E\in\mathbb{R}$ and spatial momentum $\mathbf{p}\in\APS^1$. In short, Lorentz spinors are square roots of spacetime momentum, and these square roots are simply the transformations that take the rest momentum $mc$ to $p$.

When a Lorentz spinor is Lorentz-transformed, for a Lorentz transformation $\Lambda\in\Spin(1,3)\approx\SL(2,\mathbb{C})$, the result is a Lorentz spinor at the new momentum $\overline{p}=\Lambda p\Lambda^\dagg$ times a WLG transformation. For a chiral Lorentz spinor, this gives
\begin{equation}\label{EQW: Chiral Case}
    \begin{aligned}
        \Lambda\la(p)&=\sqrt{mc}\Lambda R_pL_a \\
        &= \sqrt{mc}R_{\overline{p}}L_{\overline{a}}L_{\overline{a}}^-R_{\overline{p}}^\dagg\Lambda R_pL_a \\
        &=\la(\overline{p})L_{\overline{a}}^-R_{\overline{p}}^\dagg\Lambda R_pL_a = \la(\overline{p})S(p,\Lambda),
    \end{aligned}
\end{equation}
where $\la(\overline{p})=\sqrt{mc}R_{\overline{p}}L_{\overline{a}}$ is the Lorentz spinor for the boost $L_{\overline{a}}$ with \textit{new} rapidity $\overline{\eta}$ (still in the $\hat{a}$ direction) and the rotor $R_{\overline{p}}$ that takes $\hat{a}$ to $\hat{\overline{p}}$. The WLG transformation is identified as
\begin{equation}\label{EQW: Chiral WLG}
    S(p,\Lambda) = L_{\overline{a}}^-R_{\overline{p}}^\dagg\Lambda R_p L_a.
\end{equation}
It can be easily demonstrated that both $SS^\dagg=S^\dagg S=1$ and $S\widetilde{S}=\widetilde{S}S=1$, so $S\in\Spin(3)\approx\SU(2)$. Indeed, this is the WLG for massive particles. There exists an identical finding for achiral Lorentz spinors. Keeping arguments implicit,
\begin{equation}\label{EQW: Achiral Case}
    \begin{aligned}
        \Lambda\la_p&=\sqrt{mc}\Lambda L_p \\
        &= \sqrt{mc}L_{\overline{p}}L_{\overline{p}}^-\Lambda L_p\\
        &=\la_{\overline{p}}L_{\overline{p}}^-\Lambda L_p = \la_{\overline{p}}S_\Lambda,
    \end{aligned}
\end{equation}
where $\la_{\overline{p}}=\sqrt{mc}L_{\overline{p}}$ is the achiral Lorentz spinor at the new momentum. The WLG transformation in the case of an achiral spinor is written with a $\Lambda$ subscript to distinguish it from the chiral case. It is identified as
\begin{equation}\label{EQW: Achiral WLG}
    S_\Lambda = L_{\overline{p}}^-\Lambda L_p = R_{\overline{p}}SR_p^\dagg.
\end{equation}
Akin to the chiral case, it can be easily shown that $S_\Lambda\in\Spin(3)$. Again, this is the WLG for massive particles.

\subsection{Massless Lorentz Spinors}

Massless Lorentz spinors are simply the massless limits ($\lim_{m\to0}\la$) of massive Lorentz spinors. Thus a massless chiral Lorentz spinor is found to be
\begin{equation}\label{EQW: Massless Chiral Spinor}
    \la_{\Fo}= \sqrt{\frac{2E}{c}}R_p\ell_+,
\end{equation}
where $\ell_\pm=(1\pm\hat{a})/2$ is a lightray in the $\pm\hat{a}$ direction. This massless spinor is a forward-oriented \textit{ray spinor} as defined in SEC.~\ref{B}. A massless achiral Lorentz spinor is then
\begin{equation}\label{EQW: Massless Achiral Spinor}
    \la_{+p}=\la_{\Fo} R^\dagg_p=\sqrt{\frac{2E}{c}}p_+,
\end{equation}
where $p_\pm=(1\pm\hat{p})/2$ is a lightray in the $\pm\hat{p}$ direction. Akin to massive spinors, these massless spinors satisfy the definition of \textit{massless} spacetime momentum,
\begin{equation}\label{EQW: Massless Momentum}
    p = \frac{2E}{c}p_+=\frac{E}{c}(1+\hat{p}) = \la_\Fo\la_\Fo^\dagg = \la_{+p}\la_{+p}^\dagg.
\end{equation}
Lorentz transformations of this lightlike momentum $p$, when changing from a reference frame with energy $E$ to a frame with energy $\overline{E}$, have rapidity $\eta = \ln{(\overline{E}/E)}$ and so can be decomposed as
\begin{equation}\label{EQW: Massless LT}
    \Lambda|_{m=0} = RL|_{m=0} = Re^{\frac{1}{2}(\eta|_{m=0}) \hat{p}}=Re^{\frac{1}{2}\ln{\frac{\overline{E}}{E}}\hat{p}}.
\end{equation}
When a massless Lorentz spinor is transformed by this Lorentz rotor, the result is a massless Lorentz spinor at the new momentum times a WLG transformation. For a chiral Lorentz spinor, this is
\begin{equation}\label{EQW: Massless chiral case}
    \begin{aligned}
        (\Lambda|_{m=0})\la_{\Fo}(p)&=\sqrt{\frac{2E}{c}}\Lambda R_p\ell_+ \\
        &= \sqrt{\frac{2E}{c}}Re^{\frac{1}{2}\ln{\frac{\overline{E}}{E}}\hat{p}}R_p\ell_+\\
        &=\sqrt{\frac{2E}{c}}Re^{\frac{1}{2}\ln{\frac{\overline{E}}{E}}\hat{p}}p_+R_p \\
        &=\sqrt{\frac{2E}{c}}Re^{\frac{1}{2}\ln{\frac{\overline{E}}{E}}}p_+R_p\\
        &=\sqrt{\frac{2E}{c}}\sqrt{\frac{\overline{E}}{E}}R_{\overline{p}}R_{\overline{p}}^\dagg RR_p\ell_+ \\
        &=\sqrt{\frac{2\overline{E}}{c}}R_{\overline{p}}e^{-i\frac{1}{2}\omega\hat{a}}\ell_+ = e^{-i\frac{1}{2}\omega}\la_\Fo(\overline{p}),
    \end{aligned}
\end{equation}
where $\la_\Fo(\overline{p})=\sqrt{2\overline{E}/c}\,R_{\overline{p}}\ell_+$ is the Lorentz spinor with \textit{new} energy $\overline{E}\in\mathbb{R}$ and $\omega=\omega(p,\Lambda)\in\mathbb{R}$ is the WLG phase. For an achiral Lorentz spinor, keeping arguments implicit, and forgoing the full derivation as it follows from EQ.~\ref{EQW: Massless chiral case},
\begin{equation}\label{EQW: Massless achiral case}
     (\Lambda|_{m=0})\la_{+p}=\lambda_{+\overline{p}}R_{\overline{p}}e^{-i\frac{1}{2}\omega\hat{a}}R_p^\dagg=e^{-i\frac{1}{2}\omega}\lambda_{+\overline{p}}R_{\overline{p}}R_p^\dagg,
\end{equation}
where $\la_{+\overline{p}}=\sqrt{2\overline{E}/c}\,\overline{p}_+$. The rotor factor $R_{\overline{p}}R_p^\dagg$ might seem like a feature unique to the massless case, but it is simply because of the relationship between WLG transformations in the chiral and achiral case. This may be alternately confirmed by considering the massless limits of the massive WLG transformations, $S$ and $S_\Lambda$ (EQ.~\ref{EQW: Chiral WLG} and EQ.~\ref{EQW: Achiral WLG}). For the former, using some reference energy $E_0\in\BR$,
\begin{equation}\label{EQW: Massless limit of chiral WLG}
    \begin{aligned}
        S|_{m=0} &= (L^-_{\overline{a}}|_{m=0})R_{\overline{p}}^\dagg R (\Lambda|_{m=0})R_p(L_a|_{m=0}) \\
        &= \sqrt{\frac{E_0}{\overline{E}}}\ell_+R_{\overline{p}}^\dagg R\sqrt{\frac{\overline{E}}{E}}p_+R_p\sqrt{\frac{E}{E_0}}\ell_+ \\
        &= \ell_+e^{-i\frac{1}{2}\omega\hat{a}}\ell_+ = e^{-i\frac{1}{2}\omega}\ell_+.
    \end{aligned}
\end{equation}
And for the latter, the calculation follows from the relation $S_\Lambda=R_{\overline{p}}SR_p^\dagg$:
\begin{equation}\label{EQW: Massless limit of achiral WLG}
    \begin{aligned}
        S_\Lambda|_{m=0} &= R_{\overline{p}}(S|_{m=0})R_p^\dagg \\
        &= e^{-i\frac{1}{2}\omega}R_{\overline{p}}\ell_+R_p^\dagg = e^{-i\frac{1}{2}\omega}\overline{p}_+R_{\overline{p}}R_p^\dagg.
    \end{aligned}
\end{equation}
The above derivations show that the key WLG transformation at play is $e^{-i\omega/2}$. This is an element of $\Spin(2)\approx\mathrm{U}(1)$, which is the \textit{physically realizable} subgroup of the WLG for massless particles, $\Spin(2,0,1)/\{\pm1\}\approx\mathrm{SE}(2)$ (the symmetry group of $2$-dimensional Euclidean space). All other transformations within the WLG for massless particles are \textit{lightlike} (translational) \textit{gauge transformations} and so are not physically realizable \cite{Croft2025,Han1981,Han1982,Kim2016}.  

\subsection{Scattering Algebra Lorentz Spinors}

All traditional studies of spinors in the CSM involved only the chiral Lorentz spinors of EQ.~\ref{EQW: Chiral Lorentz spinor} and EQ.~\ref{EQW: Massless Chiral Spinor}. As such, this subsection will first inject the chiral Lorentz spinors into their SA brackets--defined by the \textit{lightray filter} of EQ.~\ref{EQB: Redef. of SA Bracket}. Thereafter, the achiral Lorentz spinors of EQ.~\ref{EQW: Achiral Lorentz spinor} and EQ.~\ref{EQW: Massless Achiral Spinor} will be injected into their SA brackets.

The massive chiral Lorentz spinor of EQ.~\ref{EQW: Chiral Lorentz spinor} has an SA bracket that is equivalent to the CSM's \textit{right-angle} spin spinor:
\begin{equation}\label{EQW: Massive Chiral SA Bracket}
    \SAb{\la(p)}_{\al}^{\ \jj}\quad\leftrightarrow\quad |\mathbf{j}(p)\rangle_\al^{\ \jj}.
\end{equation}
For the SA bracket, the indices $\al,\jj\in\{+,-\}$ indicate lightray-alignment as shown in SEC.~\ref{B}. For the spin spinor, $\al\in\{1,2\}$ is the \textit{Lorentz-index} (row-index) and $\jj\in\{1,2\}$ is the \textit{spin-index} (column-index). The injection of this Lorentz spinor into its SA bracket naturally results in an SA equivalent to EQ.~\ref{EQW: Chiral Case},
\begin{equation}\label{EQW: Massive SA Chiral Case}
    \SAb{\Lambda}_{\al}^{\ \be}\SAb{\la(p)}_{\be}^{\ \jj}=\SAb{\la(\overline{p})}_{\al}^{\ \kk}\SAb{S}_{\kk}^{\ \jj}.
\end{equation} 
This is identical to EQ.~$\text{A140}$ of \cite{Christensen2024}. The massless chiral Lorentz spinor of EQ.~\ref{EQW: Massless Chiral Spinor} has an SA bracket that is equivalent to the CSM's \textit{right-angle} helicity spinor:
\begin{equation}\label{EQW: Massless Chiral SA Bracket}
    \SAb{\la_{\Fo}(p)}_{\al}^{\ +}\quad\leftrightarrow\quad |j(p)\rangle_\al.
\end{equation}
Notice that the SA bracket version has a constant index $\jj = +$. Thus the SA equivalent to EQ.~\ref{EQW: Massless chiral case} is
\begin{equation}\label{EQW: Massless SA Chiral Case}
    \SAb{\Lambda}_{\al}^{\ \be}\SAb{\la_{\Fo}(p)}_{\be}^{\ +}=e^{-i\frac{1}{2}\omega}\SAb{\la_{\Fo}(\overline{p})}_{\al}^{\ +}.
\end{equation}
This equation is similarly identical to EQ.~$\text{A138}$ of \cite{Christensen2024}.

The achiral Lorentz spinors have no identical forms in any traditional studies of the CSM, thus the SA equivalents of EQ.~\ref{EQW: Achiral Case} and EQ.~\ref{EQW: Massless achiral case} are novel with respect to the CSM. The massive achiral Lorentz spinor of EQ.~\ref{EQW: Achiral Lorentz spinor} has the SA bracket
\begin{equation}\label{EQW: Massive Achiral SA Bracket}
    \SAb{\la_p}_{\al}^{\ \jj} = \SAb{\la(p)}_{\al}^{\ \kk}\SAb{R_p^\dagg}_{\kk}^{\ \jj}.
\end{equation}
The SA equivalent of EQ.~\ref{EQW: Achiral Case} is then
\begin{equation}\label{EQW: Massive SA achiral Case}
    \SAb{\Lambda}_{\al}^{\ \be}\SAb{\la_p}_{\be}^{\ \jj}=\SAb{\la_{\overline{p}}}_{\al}^{\ \kk}\SAb{S_\Lambda}_{\kk}^{\ \jj}.
\end{equation}
The massless achiral Lorentz spinor of EQ.~\ref{EQW: Massless Achiral Spinor} likewise has the SA bracket
\begin{equation}\label{EQW: Massless Achiral SA Bracket}
    \SAb{\la_{+p}}_{\al}^{\ \jj} = \SAb{\la_{\Fo}(p)}_{\al}^{\ +}\SAb{R_p^\dagg}_{+}^{\ \jj}.
\end{equation}
So the SA equivalent of EQ.~\ref{EQW: Massless achiral case} is
\begin{equation}\label{EQW: Massless SA achiral Case}
    \SAb{\Lambda}_{\al}^{\ \be}\SAb{\la_{+p}}_{\be}^{\ \jj}=e^{-i\frac{1}{2}\omega}\SAb{\la_{+\overline{p}}}_{\al}^{\ \kk}\SAb{R_{\overline{p}}R_p^\dagg}_{\kk}^{\ \jj}.
\end{equation}
Unlike the chiral case, the SA bracket of a massless achiral Lorentz spinor has two indices. This could shed light on why the achiral Lorentz spinors were never discovered in the traditional methods for the CSM: The SA bracket of a massless achiral Lorentz spinor lacks the convenience of a single index. 

\subsection{Lorentz Products}

As detailed in \cite{Croft2026}, for $\la_j=\la(p_j)$, the SA approach to the CSM makes use of the \textit{Lorentz}(-invariant) \textit{products}
\begin{equation}\label{EQW: Lorentz Products}
    \LP{j}{k}\quad\text{and}\quad\mLP{j}{k},
\end{equation}
which are respectively left- and right-chiral, and where $\hat{B}=\hat{b}\hat{a}\in\APS^2$ is the unit bivector\footnote{This bivector is discussed in SEC.~\ref{F} and SEC.~\ref{B}.} such that $\hat{B}\hat{a}\hat{B}^\dagg=-\hat{a}$. It is therefore worthwhile to examine the WLG transformations of these products; first within the APS and then the SA. The results are of the same form whether chiral or achiral Lorentz spinors are used, so this subsection will take (the massive spinor) $\la_j$ to be chirality-agnostic with an appropriate WLG transformation, $\mathcal{S}_j$. 

The left-chiral Lorentz product is Wigner-covariant,
\begin{equation}\label{EQW: Left LP Covariance}
    \hat{B}\widetilde{\Lambda\la_j}\Lambda\la_k = (\hat{B}\mathcal{S}_j^\dagg\hat{B}^\dagg)\hat{B}\widetilde{\la}_{\overline{j}}\la_{\overline{k}}\mathcal{S}_k,
\end{equation}
where $\widetilde{\la}_{\overline{j}}$ and $\la_{\overline{k}}$ are the Lorentz spinors at their new momenta. The right-chiral Lorentz product is identically Wigner-covariant,
\begin{equation}\label{EQW: Right LP Covariance}
    \begin{aligned}
        \hat{B}^\dagg(\Lambda\la_j)^\dagg\Lambda^-\la^-_k &= (\hat{B}^\dagg\mathcal{S}_j^\dagg\hat{B})\hat{B}^\dagg\la^\dagg_{\overline{j}}\la^-_{\overline{k}}\mathcal{S}_k \\
        &=(\hat{B}\mathcal{S}_j^\dagg\hat{B}^\dagg)\hat{B}^\dagg\la^\dagg_{\overline{j}}\la^-_{\overline{k}}\mathcal{S}_k,
    \end{aligned}
\end{equation}
because the WLG transformations are identical to the left-chiral product's. The transformations to the left of the spinors are sandwiched between the bivector $\hat{B}$; this will require multiplication, by the bivector, of the constructive fields in SEC.~\ref{F}. Without this multiplication, the Lagrangian density terms would contain extraneous factors of $\hat{B}$. 

Applying the lightray filter of EQ.~\ref{EQB: SA Bracket}, the SA equivalent of EQ.~
\ref{EQW: Left LP Covariance} is
\begin{equation}\label{EQW: Left LP in SA}
    \begin{aligned}
         \upSAb{\hat{B}\widetilde{\Lambda\la_j}\Lambda\la_k}{\jj}{\kk} &= \SAb{\hat{B}\mathcal{S}_j^\dagg\hat{B}^\dagg}^{\jj}_{\ \Le}\upSAb{\LP{\overline{j}}{\overline{k}}}{\Le}{\mm}\SAb{\mathcal{S}_k}_{\mm}^{\ \kk}, \\
    \end{aligned}
\end{equation}
and the SA equivalent of EQ.~\ref{EQW: Right LP Covariance} is
\begin{equation}\label{EQW: Right LP in SA}
    \begin{aligned}
        \upSAb{\hat{B}^\dagg(\Lambda\la_j)^\dagg\Lambda^-\la^-_k}{\jj}{\kk} &= \SAb{\hat{B}\mathcal{S}_j^\dagg\hat{B}^\dagg}^{\jj}_{\ \Le}\upSAb{\mLP{\overline{j}}{\overline{k}}}{\Le}{\mm}\SAb{\mathcal{S}_k}_{\mm}^{\ \kk}. \\
    \end{aligned}
\end{equation}
These equations correspond to the transformations discussed throughout the appendix of \cite{Christensen2024}.

In the massless limit, and for the moment only considering chiral Lorentz spinors, the Lorentz products of EQ.~\ref{EQW: Lorentz Products} become
\begin{equation}\label{EQW: Massless Lorentz Products}
    \hat{B}\widetilde{\la_{j\Fo}}\la_{k\Fo}=-\la^\Fo_j\la_{k\Fo}\quad\text{and}\quad\hat{B}^\dagg(\la_{j\Fo})^\dagg\la_{k\Fo}^-=-\la^\Ba_j\la_{k\Ba},
\end{equation}
where $j$ and $k$ are still particle labels and the Lorentz products have been simplified into the symplectic form of EQ.~\ref{EQB: Symplectic} using the methods detailed in \cite{Croft2026} and briefly summarized in SEC.~\ref{B}. Explicitly, recalling that $\hat{B}=\hat{b}\hat{a}$, $\la_{\Ba}=\la_{\Fo}^-$ so $\la^{\Fo/\Ba}=\hat{b}\widetilde{\la_{\Fo/\Ba}}$. These Lorentz products respectively map to $\mathrm{Center}[\APS]\ell_+=\APS^{0\oplus3}\ell_+$ and $\mathrm{Center}[\APS]\ell_-=\APS^{0\oplus3}\ell_-$ which are both isomorphic to $\mathbb{C}$. Thus, when injected into their SA brackets,
\begin{equation}\label{EQW: Massless SA Brack of LP}
    \upSAb{\la^\Fo_j\la_{k\Fo}}{+}{+}\quad\text{and}\quad\upSAb{\la^\Ba_j\la_{k\Ba}}{-}{-},
\end{equation}
they must be elements of $\mathrm{Center}[\APS]=\APS^{0\oplus3}\approx\mathbb{C}$: That is, they are complex numbers just like the massless products of \cite{Christensen2024}. These massless Lorentz products are inversely Wigner-covariant with respect to one another, as demonstrated by
\begin{equation}\label{EQW: WLG of massless Left LP}
    \la^\Fo_j\widetilde{\Lambda}\Lambda\la_{k\Fo} = e^{-i\frac{1}{2}(\omega_j+\omega_k)}\la^\Fo_{\overline{j}}\la_{\overline{k}\Fo}
\end{equation}
and
\begin{equation}\label{EQW: WLG of massless Left LP}
    \la^\Ba_j\widetilde{\Lambda}\Lambda\la_{k\Ba} = e^{i\frac{1}{2}(\omega_j+\omega_k)}\la^\Ba_{\overline{j}}\la_{\overline{k}\Ba}.
\end{equation}
Their SA equivalents are then
\begin{equation}\label{EQW: SA WLG massless left LP}
    \upSAb{\la^\Fo_j\widetilde{\Lambda}\Lambda\la_{k\Fo}}{+}{+}=e^{-i\frac{1}{2}(\omega_j+\omega_k)}\upSAb{\la^\Fo_{\overline{j}}\la_{\overline{k}\Fo}}{+}{+}
\end{equation}
and
\begin{equation}\label{EQW: SA WLG massless right LP}
    \upSAb{\la^\Ba_j\widetilde{\Lambda}\Lambda\la_{k\Ba}}{-}{-}=e^{i\frac{1}{2}(\omega_j+\omega_k)}\upSAb{\la^\Ba_{\overline{j}}\la_{\overline{k}\Ba}}{-}{-}.
\end{equation}
These correspond to the massless WLG transformations similarly discussed throughout the appendix of \cite{Christensen2024}. 

The situation looks much more complicated for the achiral case, yet it is but an illusion. Once more taking the massless limit, now considering achiral Lorentz spinors, the Lorentz products of EQ.~\ref{EQW: Lorentz Products} become
\begin{equation}\label{EQW: Achiral Massless left LP}
    \hat{B}\widetilde{\la_{+p_j}}\la_{+p_k} = -\hat{B}R_{p_j}\hat{B^\dagg}\la_j^\Fo\la_{k\Fo}R_{p_k}^\dagg
\end{equation}
and
\begin{equation}\label{EQW: Achiral Massless right LP}
    \hat{B}^\dagg\la_{+p_j}^\dagg\la_{+p_k}^- = -\hat{B}R_{p_j}\hat{B^\dagg}\la_j^\Ba\la_{k\Ba}R_{p_k}^\dagg.
\end{equation}
Despite their prickly appearance, these Lorentz products will maintain Wigner-covariance due to their explicit dependence upon $\la_j^\Fo\la_{k\Fo}$ and $\la_j^\Ba\la_{k\Ba}$. Again, the apparently---but nonetheless illusorily---complicated nature of achiral spinors serves as a potential reason for why they were never discovered using the traditional methods for the CSM. 

As the above products are so messy, and as there is no equivalent in the literature, the SA equivalents for EQ.~\ref{EQW: Achiral Massless left LP} and EQ.~\ref{EQW: Achiral Massless right LP} will be ignored in this paper. A further reason is that when inside Lagrangian density terms, which pair Lorentz spinors with appropriately\footnote{It will always be possible to factor the Lagrangian density terms into either products of achiral Lorentz spinors and to-be-introduced \textit{unprimed} fields, or products of chiral Lorentz spinors and to-be-introduced \textit{primed} fields.} and inversely Wigner-covariant fields, all results are chirality-agnostic. This extends the chirality-agnostic results of \cite{Croft2026}.

\section{Constructive Spin Fields}\label{F}

Lagrangian density terms in the Constructive Standard Model (CSM) involve a contraction over indices that results in a trace of products of constructive spin fields with spin/helicity spinors \cite{Christensen2024}. By SEC.~\ref{B}, summarizing \cite{Croft2026}, this will correspond to the \textit{central projection} of products of spin fields with Lorentz spinors. In order for the action to be Lorentz-invariant, the spin fields must transform \textit{inversely} to the Lorentz rotors of SEC.~\ref{W}. That is, \textit{constructive} spin fields must be inversely Wigner-covariant with respect to Lorentz spinors.  

Using the APS, this section will first introduce spin fields and density elements following the methods of \cite{McKenzie2015} and \cite{Burns2025}. It will be seen how (normalized) spin-$1/2$ fields are simply rotors, and that (normalized) density elements are then rotations of the lightray $\ell_+$ by said rotors. Spin-$1$ fields will then be a logical consequence of letting the density element's \textit{interference} term be independent. Following the structure of SEC.~\ref{W}, this section will end by injecting the objects from the APS into the Scattering Algebra (SA). The methods of this section serve as a novel first-bridge between studies of the CSM and of Quantum Information.

\subsection{Massive Spin Fields and Density}

Assuming the notation of SEC.~\ref{B}---first established in \cite{Croft2026}---a massive spin-$1/2$ field may be defined, without loss of generality, as 
\begin{equation}\label{EQF: Spin-1/2 Field}
    f (p)= f_\uparrow(p)-\hat{B}f_\downarrow(p).
\end{equation}
The coefficients $f_{\uparrow/\downarrow}\in\mathbb{R}$ are the field's respective\footnote{Within the frame induced by momentum $p$.} \textit{participation-amplitudes} for spin-$\uparrow$ and spin-$\downarrow$, and $\hat{B}=\hat{b}\hat{a}\in\APS^2$ is the unit bivector satisfying the \textit{flip-rotation} $\hat{B}\hat{a}\hat{B}^\dagg=-\hat{a}$. To be precise, $f$ holds the \textit{spin with respect to} $\hat{a}$. This paper uses the term participation-amplitude to highlight the interpretation developed in \cite{McKenzie2015}: That $f_\uparrow=|f|\sqrt{P_\uparrow}$ denotes the strength-amplitude $|f|$ times the probability-amplitude $\sqrt{P_\uparrow}$ of the \textit{identity-rotation}'s participation in the transformation encoded within the field, and likewise that $f_\downarrow=|f|\sqrt{P_\downarrow}$ denotes the same information but for the flip-rotation's participation. The field density element (FDE) is defined, with implicit arguments, as the application of $f$ to the lightray $\ell_+$,
\begin{equation}\label{EQF: Field Density Element}
\begin{aligned}
    \rho = f\ell_+ f^\dagg &= |f_\uparrow|^2\ell_+ + |f_\downarrow|^2\ell_- - |f_\mathrm{int}|^2\hat{b} \\
    &= |f|^2\left(P_\uparrow\ell_++P_\downarrow\ell_--|P_\mathrm{int}|\hat{b}\right),
\end{aligned}
\end{equation}
where $|f_\mathrm{int}|=|f|^2|P_\mathrm{int}|=f_\uparrow f_\downarrow$ is the interference term of the FDE and $P_\uparrow+P_\downarrow=1$. Clearly, the usual spin density element (DE) can be recovered from the FDE:
\begin{equation}\label{EQF: Spin Density Element}
    \rho_s = \rho/|f|^2 = P_\uparrow\ell_++P_\downarrow\ell_--|P_\mathrm{int}|\hat{b}.
\end{equation}
This corresponds to normalizing the field in EQ.~\ref{EQF: Spin-1/2 Field},
\begin{equation}\label{EQF: Spin Rotor}
    R_s =f/|f|= \sqrt{P_\uparrow}-\hat{B}\sqrt{P_\downarrow},
\end{equation}
which is a rotor satisfying the definition of the \textit{operatorial Pauli spinor} (spin rotor with respect to $\hat{a}$) in EQ.~\ref{EQB: R_s}. That is, $R_s\hat{a}R_s^\dagg=\hat{s}\in\APS^1$ is the \textit{spin vector}. This implies that the DE can be rewritten as
\begin{equation}\label{EQF: Spin Density Element 2}
    \rho_s = R_s\ell_+R_s^\dagg = \frac{1}{2}(1+\hat{s}).
\end{equation}
Clearly, its geometric interpretation is that of a lightray in the direction of $\hat{s}$. 

The spin-$1/2$ field and FDE were presented first to show that when normalized, the traditional objects---Pauli spinors and DEs---naturally fall out. This therefore serves as a bridge between the CSM and the discipline of Quantum Information. 

The discussion of spin-$1$ fields is now at hand. Looking qualitatively at the form of the FDE in EQ.~\ref{EQF: Field Density Element}, the interference term conceptually behaves as if it were spin-$0$. It is important to understand, however, that it is \textit{not} a spin-$0$ term---the same as saying that density elements do not represent spin-$1$ objects. This is because the FDE is able to be written as a spin-$1/2$ field \textit{transforming} a lightray. In traditional Physics-speak: Spin-$1$ fields cannot be factored into the square of a spin-$1/2$ field. Therefore, the logical choice is to just allow the coefficient of $\hat{b}$ within the FDE to be independent of the other coefficients. This results in the definition of a (massive) spin-$1$ field,
\begin{equation}\label{EQF: Spin-1 Field}
    Z(p) = Z_\uparrow(p)\ell_++Z_\downarrow(p)\ell_-+Z_0(p)\hat{b}.
\end{equation}
The coefficients $Z_{\uparrow/\downarrow/0}\in\mathbb{R}$ represent the probability-amplitudes of measuring the polarization of $Z$ as spin-$\uparrow$, spin-$\downarrow$, and spin-$0$.

\subsection{Massless Spin Fields}

A massless spin-$1/2$ particle has either \textit{helicity}-$\uparrow$ or \textit{helicity}-$\downarrow$, with no mixing. For the helicity-$\uparrow$ case the field is
\begin{equation}\label{EQF: Massless Spin-Up Spin-Half}
    \nu_-(p) = \nu_\uparrow(p),
\end{equation}
and for the helicity-$\downarrow$ case the field is
\begin{equation}\label{EQF: Massless Spin-Down Spin-Half}
    \nu_+(p) = -\hat{B}\nu_\downarrow(p).
\end{equation}
The density elements of EQ.~\ref{EQF: Field Density Element} and EQ.~\ref{EQF: Spin Density Element} are of little practical use in the context of massless fields, as there is no mixing. Hence their discussion in a massless context will be forgone. Yet, spin-$1$ fields continue to be practical despite the ignoring of density elements. In fact, spin-$1$ fields polarize \textit{fully} to either helicity-$\uparrow$ or helicity-$\downarrow$ just like spin-$1/2$ fields. For the helicity-$\uparrow$ case,
\begin{equation}\label{EQF: Massless Spin-Up Spin-One}
    A_-(p) = A_\uparrow(p)\ell_+,
\end{equation}
while for the helicity-$\downarrow$ case,
\begin{equation}\label{EQF: Massless Spin-Down Spin-One}
    A_+(p) = A_\downarrow(p)\ell_-.
\end{equation}
While the massless fields lose mixing and therefore the degrees of freedom inherent to massive fields, their interpretations follow from the massive case.

\subsection{Wigner Little Groups and Spin Fields}

 Constructive fields do not transform under the Lorentz group. Instead, they transform under their appropriate Wigner little groups (WLGs) such that they are \textit{inversely} Wigner-covariant with respect to Lorentz spinors. For massive spin-$1/2$ fields, the WLG is $\Spin(3)\approx\SU(2)$. For massless spin-$1/2$ fields, the physically realizable subgroup of its WLG is $\Spin(2)\approx\mathrm{U}(1)\subset\Spin(2,0,1)\approx\pm\mathrm{SE}(2)$.

Following the methods of WLG-based field theory presented in \cite{Christensen2024}---and rewriting them in the language of maps from $\APS$ to $\APS$---the WLG transformation of the massive spin-$1/2$ field is
\begin{equation}\label{EQF: WLG of f}
    f(p)\mapsto S^\dagg_\Lambda f(\overline{p})=S^\dagg_\Lambda\left(f_\uparrow(\overline{p})-\hat{B}f_\downarrow(\overline{p})\right),
\end{equation}
where $S_\Lambda^\dagg\in\Spin(3)$ is the reverse (Hermitian conjugate) of the achiral WLG transformation in EQ.~\ref{EQW: Achiral Case} and $f_{\uparrow/\downarrow}(\overline{p})\in\mathbb{R}$ are the participation-amplitudes within the frame induced by momentum $\overline{p}=\Lambda p\Lambda^\dagg$. Thus, for a massive spin-$1$ field, the WLG transformation must be
\begin{equation}\label{EQF: WLG of A}
\begin{aligned}
    Z(p)\mapsto \,\, &S_\Lambda^\dagg Z(\overline{p})S_\Lambda \\
    &= S_\Lambda^\dagg\left(Z_\uparrow(\overline{p})\ell_++Z_\downarrow(\overline{p})\ell_-+Z_0(\overline{p})\hat{b}\right) S_\Lambda.
\end{aligned}
\end{equation}
These mappings give clear physical meaning: When a reference frame is changed via a Lorentz transformation, the constructive field's values change accordingly and a back-rotation (the rotor $S_\Lambda^\dagg$) is applied to account for the change in physical orientation. 

For the massless cases, and for the extent of this paper, it suffices to solely discuss the transformations of the helicity-$\uparrow$ fields. The WLG transformation of a spin-$1/2$ helicity-$\uparrow$ field is therefore
\begin{equation}\label{EQF: WLG of nu}
    \nu_{\uparrow}(p)\mapsto R_p e^{i\frac{1}{2}\omega\hat{a}}R_{\overline{p}}^\dagg\nu_{\uparrow}(\overline{p}),
\end{equation}
where $R_pe^{i\omega\hat{a}/2}R_{\overline{p}}^\dagg$ is the inverse of the WLG transformation in EQ.~\ref{EQW: Massless achiral case}. For a spin-$1$ helicity-$\uparrow$ field, the WLG transformation is
\begin{equation}\label{EQF: WLG of massless A}
        A_\uparrow(p)\ell_+\mapsto R_pe^{i\frac{1}{2}\omega\hat{a}}R_{\overline{p}}^\dagg A_\uparrow(\overline{p})\ell_+R_{\overline{p}}e^{-i\frac{1}{2}\omega\hat{a}}R_p^\dagg
\end{equation}
Another curious thing is how the fields all transform under the WLG for \textit{achiral} Lorentz spinors. This implies the existence of pre-backrotated fields which will pair with \textit{chiral} Lorentz spinors. By virtue of being pre-backrotated, they are primed for use alongside chiral Lorentz spinors. Such fields are henceforth called \textit{primed fields}. A further implication is that any Lagrangian density terms, which are Lorentz spinors combined with fields, will be \textit{chirality-agnostic}.\footnote{Please note that Lorentz products---representing dynamical interactions---bear chirality regardless of the Lorentz spinors' chirality and so "chirality-agnostic" does not mean to imply there are no chiral interactions.} A massive spin-$1/2$ \textit{primed} field is of the form
\begin{equation}\label{EQF: Primed massive spin-half field}
    f'(p)=R_p^\dagg f(p),
\end{equation}
and its WLG transformation is
\begin{equation}\label{EQF: WLG of primed massive spin-half}
    f'(p)\mapsto S^\dagg f'(\overline{p})=S^\dagg R_{\overline{p}}^\dagg f(\overline{p}).
\end{equation}
A massive spin-$1$ \textit{primed} field is of the form
\begin{equation}\label{EQF: Primed massive spin-one field}
    Z'(p)=R_p^\dagg Z(p)R_p,
\end{equation}
and its WLG transformation is
\begin{equation}\label{EQF: WLG of primed massive spin-one}
    Z'(p) \mapsto S^\dagg Z'(\overline{p})S = S^\dagg R_{\overline{p}}^\dagg Z(\overline{p})R_{\overline{p}}S.
\end{equation}
The WLG transformations of these equations---$S,S^\dagg\in\Spin(3)$---are the transformation of EQ.~\ref{EQW: Chiral Case} and its Hermitian conjugate. 

Primed fields exist in the massless cases as well. The spin-$1/2$ \textit{primed} helicity-$\uparrow$ field is of the same form as EQ.~\ref{EQF: Primed massive spin-half field} and its WLG transformation is 
\begin{equation}\label{EQF: WLG of primed massless spin-half}
    \begin{aligned}
        \nu'_-(p)=\nu_-(p)R_p^\dagg&\mapsto e^{i\frac{1}{2}\omega\hat{a}}\nu_-(\overline{p})R_{\overline{p}}^\dagg\\
        &=e^{i\frac{1}{2}\omega\hat{a}}\nu'_-(\overline{p}).
    \end{aligned}
\end{equation}
Lastly, the spin-$1$ \textit{primed} helicity-$\uparrow$ field is of the same form as EQ.~\ref{EQF: Primed massive spin-one field} and its WLG transformation is
\begin{equation}\label{EQF: WLG of primed massless spin-one}
    \begin{aligned}
        A'_-(p)=A_\uparrow(p)R_p^\dagg\ell_+R_p&\mapsto e^{i\frac{1}{2}\omega\hat{a}}A_\uparrow(\overline{p})R^\dagg_{\overline{p}}\ell_+R_{\overline{p}} e^{-i\frac{1}{2}\omega\hat{a}}\\
        &=e^{i\frac{1}{2}\omega\hat{a}} A'_-(\overline{p}) e^{-i\frac{1}{2}\omega\hat{a}}.
    \end{aligned}
\end{equation}

\subsection{Scattering Algebra Spin Fields}

The results of this subsection are of the same form regardless of a massive field's \textit{primedness} (backrotatedness), so this subsection will take all massive fields to be \textit{primedness-agnostic} with an appropriate WLG transformation of $\mathcal{S}$.

Placing constructive spin fields within the SA necessitates delicate handling. Firstly and for convenience, spin fields are injected into SA brackets with \textit{fully lowered} indices and therefore require the lightray filter
\begin{equation}\label{EQF: Lowered SA Bracket}
    \downSAb{g}{\kk}{\jj}=\ell_\jj g\ell_\kk + \hat{b}\ell_\jj g\ell_\kk\hat{b}
\end{equation}
as opposed to that of EQ.~\ref{EQB: SA Bracket} or EQ.~\ref{EQB: Redef. of SA Bracket}. Secondly, the end of SEC.~\ref{W} showed how Lorentz products contain WLG transformations sandwiched between the bivector $\hat{B}$---thereby requiring the rightwise multiplication of fields by said bivector. Thirdly, spin-$1$ fields will lose index-symmetry when multiplied by $\hat{B}$ and therefore are in seeming conflict with the constructive fields of \cite{Christensen2024}. In reality, the lack of index-symmetry is required by the pseudoreal nature of $\Spin(3)\approx\SU(2)$. Fourthly, massive spin-$1/2$ fields in the CSM have only \textit{one} index due to the fields' two degrees of freedom; meaning that fields within the SA must be multiplied by a lightray---$\ell_+$ by convention---to reduce the index-count to one. There are two ways to do this: For an arbitrary multivector, $g\in\APS$, the first is demonstrated by
\begin{equation}\label{EQF: Reduced Index Property}
    \downSAb{g\ell_+}{\kk}{\jj}=\downSAb{g\ell_+}{+}{\jj}=\downSAb{g}{+}{\jj}
\end{equation}
and the second by
\begin{equation}\label{EQF: Reduced Index Property 2}
    \downSAb{\ell_+g}{\kk}{\jj}=\downSAb{\ell_+g}{\kk}{+}=\downSAb{g}{\kk}{+}.
\end{equation}
Either of these forms is used when appropriate, and what is appropriate will be made clear in SEC.~\ref{T}.

Delicately handling the massive spin-$1/2$ field of EQ.~\ref{EQF: Spin-1/2 Field} yields the spin-$1/2$ \textit{SA field},
\begin{equation}\label{EQF: SA Spin-1/2 Field}
    \downSAb{f(p)\ell_+}{+}{\jj}= \downSAb{f(p)}{+}{\jj},
\end{equation}
whose WLG transformation is
\begin{equation}\label{EQF: SA Spin-1/2 WLG}
    \downSAb{f(p)}{+}{\jj}\mapsto\SAb{\mathcal{S}^\dagg}_{\jj}^{\ \Le}\downSAb{f(\overline{p})}{+}{\Le}.
\end{equation}
This SA field is equivalent to the massive (lower-index particle) spin-$1/2$ field of \cite{Christensen2024}. Now, there is a second case following EQ.~\ref{EQF: Reduced Index Property 2} that yields another massive spin-$1/2$ SA field,
\begin{equation}\label{EQF: SA Spin-1/2 Field 2}
    \downSAb{\ell_+f^\dagg(p)\hat{B}}{\kk}{+}=\downSAb{f^\dagg(p)\hat{B}}{\kk}{+},
\end{equation}
and its WLG transformation is 
\begin{equation}\label{EQF: SA Spin-1/2 WLG 2}
    \downSAb{f^\dagg(p)\hat{B}}{\kk}{+}\mapsto\downSAb{f^\dagg(\overline{p})\hat{B}}{\mm}{+}\SAb{\hat{B}\mathcal{S}\hat{B}^\dagg}^{\mm}_{\ \kk}.
\end{equation}
This SA field is equivalent to the massive (lower-index antiparticle) spin-$1/2$ field of \cite{Christensen2024}. In some cases, when two spin-$1/2$ fields $f_1=f(p_1)$ and $f_2=f(p_2)$ are present within the same Lagrangian density, their fields will respectively be of the forms in EQ.~\ref{EQF: SA Spin-1/2 Field} and EQ.~\ref{EQF: SA Spin-1/2 Field 2} and will combine to the lone SA bracket
\begin{equation}\label{EQF: SA Spin-1/2 Field Combined}
    \downSAb{f_1}{+}{\jj}\downSAb{f_2^\dagg\hat{B}}{\kk}{+} = \downSAb{f_1\ell_+f^\dagg_2\hat{B}}{\kk}{\jj}
\end{equation}
that will be inversely Wigner-covariant with respect to Lorentz products. An example of this will be shown in the next section. 

It was mentioned that spin-$1$ fields seem unable to be represented within the SA in a manner analogous to the CSM. This is because, for the naïve massive spin-$1$ SA field,
\begin{equation}\label{EQF: SA Spin-1 Field}
    \downSAb{A(p)\hat{B}}{\kk}{\jj},
\end{equation}
index-symmetry is broken. However this is not a true problem as the spin-$1$ fields of \cite{Christensen2024} implicitly bear this structure.

Unfortunately, due to the SA's indices, it is not as easy to discuss massless terms in a primedness-agnostic way, and for the purpose of this paper only primed massless fields will be shown explicitly as SA fields: The unprimed massless fields' forms and transformations follow from the primed massless fields' forms and transformations.

Remember, any massless field of a (spin-$1/2$ or spin-$1$) particle is \textit{either} helicity-$\uparrow$ or helicity-$\downarrow$, so once again it suffices to discuss the SA fields and their transformations for helicity-$\uparrow$. The SA field of spin-$1/2$ primed helicity-$\uparrow$ field is
\begin{equation}\label{EQF: SA Massless Primed spin-half}
    \downSAb{\nu'_-(p)R_p^\dagg}{+}{\jj} = \SAb{R_p^\dagg}_{\jj}^{+}\downSAb{\nu_-(p)}{+}{+} 
\end{equation}
and its WLG transformation is
\begin{equation}\label{EQF: WLG SA Massless Primed spin-half}
    \downSAb{\nu'_-(p)R_p^\dagg}{+}{\jj} \mapsto \downSAb{e^{i\frac{1}{2}\omega\hat{a}}\nu_-'(\overline{p})R_{\overline{p}}^\dagg}{+}{\jj}. 
\end{equation}
It seems that the traditional CSM and the SA disagree heavily here, but a combination of a massless Lorentz product (EQ.~\ref{EQW: Massless Lorentz Products}) and these massless fields demonstrates Wigner-invariance:
\begin{equation}\label{EQF: Massless Wigner-Invariance Proof}
    \begin{aligned}
        &e^{-i\frac{1}{2}\left(\omega_j+\omega_k\right)}\upSAb{\la^\Fo_{\overline{j}}\la_{\overline{k}\Fo}}{+}{+}\downSAb{e^{i\frac{1}{2}\omega_k\hat{a}}R^\dagg_{\overline{p}_j}\nu_{\overline{j}-}\ell_+{\nu}^\dagg_{\overline{k}-}R_{\overline{p}_k}e^{- i\frac{1}{2}\omega_k \hat{a}}\widehat{B}}{\kk}{\jj} \\
        &= e^{-i\frac{1}{2}\left(\omega_j+\omega_k\right)}\upSAb{\la^\Fo_{\overline{j}}\la_{\overline{k}\Fo}}{+}{+}\downSAb{e^{i\frac{1}{2}\omega_k\hat{a}}R^\dagg_{\overline{p}_j}\nu_{\overline{j}-}\ell_+{\nu}^\dagg_{\overline{k}-}R_{\overline{p}_k}\widehat{B}e^{i\frac{1}{2}\omega_k \hat{a}}}{+}{+} \\
        &= e^{-i\frac{1}{2}\left(\omega_j+\omega_k\right)}\upSAb{\la^\Fo_{\overline{j}}\la_{\overline{k}\Fo}}{+}{+}e^{i\frac{1}{2}\left(\omega_j+\omega_k\right)}\downSAb{R^\dagg_{\overline{p}_j}\nu_{\overline{j}-}\ell_+{\nu}^\dagg_{\overline{k}-}R_{\overline{p}_k}\widehat{B}}{+}{+} \\
        &= \upSAb{\la^\Fo_{\overline{j}}\la_{\overline{k}\Fo}}{+}{+}\downSAb{R^\dagg_{\overline{p}_j}\nu_{\overline{j}-}\ell_+{\nu}^\dagg_{\overline{k}-}R_{\overline{p}_k}\widehat{B}}{+}{+}.
    \end{aligned}
\end{equation}
Indeed there remains equivalency between the CSM and the SA.\footnote{Of course, this calculation could have been performed much more efficiently within the central projection (equivalent to the trace).} The apparent discrepancy is that the original definitions of EQ.~\ref{EQW: Massless Lorentz Products} and EQ.~\ref{EQF: WLG SA Massless Primed spin-half} neither account for back-rotations nor describe how the scalar transformation is induced by the Lorentz product (the $\ell_+$ is induced in the second line). And for a spin-$1$ primed helicity-$\uparrow$ field, the SA field is
\begin{equation}\label{EQF: SA Massless Primed spin-one}
    \begin{aligned}
        \downSAb{A'_-(p)\hat{B}}{\kk}{\jj}&=\downSAb{A_\uparrow(p)R_p^\dagg\ell_+R_p\hat{B}}{\kk}{\jj} \\
        &= \SAb{R_p^\dagg}_{\jj}^{+}\downSAb{A_\uparrow(p)\hat{B}}{-}{+}\SAb{\hat{B}^\dagg R_p\hat{B}}^{-}_{\ \kk} 
    \end{aligned}
\end{equation}
with the WLG transformation
\begin{equation}\label{EQF: WLG SA Massless Primed spin-one}
    \begin{aligned}
       \downSAb{A'_-(p)\hat{B}}{\kk}{\jj}\mapsto\downSAb{e^{ i\frac{1}{2}\omega\hat{a}}R_{\overline{p}}^\dagg A_-(\overline{p})R_{\overline{p}}e^{- i\frac{1}{2}\omega\hat{a}}\widehat{B}}{\kk}{\jj}. 
    \end{aligned}
\end{equation}
Notice that the spin-$1/2$ primed SA field has one free index and that the spin-$1$ primed SA field has two free indices. These are not true degrees of freedom: They are leftover from the backrotations which define primed fields. 

Unfortunately, with massless spin-1 fields there are more unresolved and nontrivial problems. Namely, massless constructive fields interact with an entity known as the $x$-factor \cite{Christensen2023}, which has yet to be determined within the SA.

It is important to realize that this section's convention of lowering indices for SA brackets is arbitrary and purely visual; this allows for the Einstein summation convention. When indices are \textit{chosen} to be raised, there is no algebraic distinction from EQ.~\ref{EQF: Lowered SA Bracket}---compare with the raised definition in EQ.~\ref{EQB: SA Bracket}. More thorough discussions are contained in \cite{Croft2026} and in SEC.~\ref{B}. 

\section{Constructive Lagrangian Density Terms}\label{T}

Field-theoretic Lagrangian densities for the Constructive Standard Model (CSM) were first developed in \cite{Christensen2024} by contracting constructive spin fields over constructive vertices and free terms. Such terms are Lorentz-invariant through Wigner-covariance. This means two things: First, that the Lorentz transformation of a scattering amplitude defined at momentum $p$ is \textit{expressly equal} to a Wigner little group (WLG) transformation of the scattering amplitude defined at the new momentum $\overline{p}$. Second, that spin fields must be \textit{inversely} Wigner-covariant with respect to the scattering amplitudes. The developments of SEC.~\ref{W} covered the Wigner-covariance of the elements making up the vertices and scattering amplitudes while those of SEC.~\ref{F} covered the Wigner-covariance of spin fields. Moreover, those sections are extensions of \cite{Croft2026}, so the results of this paper are completely equivalent to those of \cite{Christensen2024}. Therefore, all Lagrangian densities from \cite{Christensen2024} can in principle be translated into the Scattering Algebra (SA) and analyzed using the geometry of the Algebra of Physical Space (APS). 

The following Lagrangian densities will be considered: A free massive/massless fermion, a free massive/massless boson, $hf\overline{f}$ (Higgs and two massive fermions), $q\overline{q}W$ (W-boson and two quarks), $f\overline{f}Z$ (Z-boson and two massive fermions), and $l\overline{\nu}W$ (W-boson, lepton, and antineutrino). The scattering amplitude for $hf\overline{f}$ will be fully expressed within the APS, in doing so becoming the first amplitude to be expressed purely within Geometric Algebra. It will be anti-Hermitian. No massless spin-$1$ terms will be considered because they contain $x$-factors \cite{Christensen2023}, which have yet to be explored using the SA.

\subsection{The $\overline{f}f$ and $WW$ Terms}

The free terms for massive fermions and the W-boson are effectively trivial; they are simply terms quadratic in the fields with attached Dirac deltas enforcing momentum conservation. For massive fermions $f_j = f(p_j)$, and using the SA field of EQ.~\ref{EQF: SA Spin-1/2 WLG}, the Lagrangian density is proportional to
\begin{equation}\label{EQT: ff term}
    \begin{aligned}
        \mathcal{L}_{ff}&\propto \delta^4(p_1+p_2)\upSAb{f^\dagg_2}{+}{\jj}\downSAb{f_1}{+}{\jj} \\
        &= \delta^4(p_1+p_2)2\cproj{\ell_+f_2^\dagg f_1\ell_+}.
    \end{aligned}
\end{equation}
This free term is simply the Pauli inner product \cite{Sobczyk2019}. For W-bosons $W_j = W(p_j)$, and using the SA field of EQ.~\ref{EQF: SA Spin-1 Field}, the Lagrangian density is proportional to
\begin{equation}\label{EQF: WW term}
    \begin{aligned}
        \mathcal{L}_{WW}&\propto \delta^4(p_1+p_2)\upSAb{\hat{B}^\dagg W_2^\dagg}{\jj}{\kk}\downSAb{W_1B}{\jj}{\kk} \\
        &= \delta^4(p_1+p_2)2\cproj{W_2^\dagg W_1}.
    \end{aligned}
\end{equation}
This free term is simply the squared magnitude of a multivector within the APS. These Lagrangian density terms have correspondingly identical terms in EQ.~$38$ of \cite{Christensen2024}.

\subsection{The $hf\overline{f}$ Term}

As this term is purely massive, the analysis will take $\la_j=\la(p_j)$ and $f_j=f(p_j)$ to respectively be chirality-agnostic and primedness-agnostic. From \cite{Christensen2024}, the $hf\overline{f}$ (Higgs and two massive fermions) Lagrangian density is linearly proportional to
\begin{equation}\label{EQT: hff term}
    \begin{aligned}
         \mathcal{L}_{hff} &\propto h(p_1)\downSAb{f_3^\dagg\hat{B}}{\kk}{+}\left(\upSAb{\LP{3}{2}}{\jj}{\kk}+\upSAb{\mLP{3}{2}}{\jj}{\kk}\right)\downSAb{f_2}{+}{\jj} \\
        &= h(p_1)\left(\upSAb{\LP{3}{2}}{\jj}{\kk}+\upSAb{\mLP{3}{2}}{\jj}{\kk}\right)\downSAb{f_2\ell_+f_3^\dagg\hat{B}}{\jj}{\kk},
    \end{aligned}
\end{equation}
where $h(p_1)\in\mathbb{R}$ is the Higgs field.

Collapsing to the center of the APS---$\mathrm{Center}[\APS]=\APS^{0\oplus3}\approx\mathbb{C}$, thereby performing the equivalent of a matrix trace---via EQ.~\ref{EQB: Central Proj. Collapse}, the Lagrangian density of EQ.~\ref{EQT: hff term} simplifies to
\begin{equation}\label{EQT: hff term simple}
    \mathcal{L}_{hf\overline{f}}\propto 2h(p_1)\cproj{\la^-_2f_2\ell_+f_3^\dagg\la_3^\dagg}-2h(p_1)\cproj{\la_2f_2\ell_+f_3^\dagg\widetilde{\la}_3}.
\end{equation}
The geometry of both terms is the same, so only one will be taken for analysis:
\begin{equation}\label{EQT: hff one term}
    2\cproj{\la_2f_2\ell_+f_3^\dagg\widetilde{\la}_3}.
\end{equation}
The spin-$1/2$ fields $f_2$ and $f_3$ are interpreted as \textit{transformations} "competing" to rotate and scale the lightray $\ell_+$. Thus $f_2\ell_+f_3^\dagg$ has an interpretation similar to the field density element of EQ.~\ref{EQF: Field Density Element}. The Lorentz spinors $\la_2$ and $\la_3$ are likewise interpreted as Lorentz boosts further "competing" to transform $f_2\ell_+f_3^\dagg$. It was shown in \cite{Croft2026} how two times the central projection, $2\cproj{\cdot}$, is equivalent to the traditional CSM's matrix-trace. This is conceptually similar to traditional Pauli Theory wherein the trace (twice the central projection) of the (spin) density element's \textit{square} in EQ.~\ref{EQF: Spin Density Element} is a measure of system purity/mixture. Geometrically, it measures the total orientation of $\la_2f_2\ell_+f_3^\dagg\widetilde{\la}_3$ that \textit{overlaps and collapses} into the timelike direction.

It is possible to explicitly relate this to Pauli Theory by making use of the known relations \cite{Sobczyk2019,Lounesto2001,Doran2003,Vaz2019}
\begin{equation}\label{EQT: Relation to Pauli Theory}
    f_j\ell_+\leftrightarrow \ket{s_j},\quad\la_2\leftrightarrow L_{p_2},\quad\text{and}\quad\widetilde{\la}_3\leftrightarrow (L_{p_3}^{-1})^\dagg,
\end{equation}
where $\ket{s_j}$ is the Pauli spinor \textit{ket} for the $j$-th fermion, and $L_{p_j}$ is the boost-matrix from rest to spacetime momentum $p_j$. Then EQ.~\ref{EQT: hff one term} can be rearranged using the central projection's cyclic property,
\begin{equation}\label{EQT: hff one term rearranged}
    2\cproj{\la_2f_2\ell_+f_3^\dagg\widetilde{\la}_3} = 2\cproj{\ell_+f_3^\dagg\widetilde{\la}_3\la_2f_2\ell_+},
\end{equation}
which is directly equivalent to the Pauli inner product:
\begin{equation}\label{EQT: Pauli Inner Product}
    2\cproj{\ell_+f_3^\dagg\widetilde{\la}_3\la_2f_2\ell_+}\leftrightarrow\bra{s_3}(L_{p_3}^{-1})^\dagg L_{p_2}\ket{s_2}.
\end{equation}
There is a pattern to be extracted from this: Applying these results to both terms in EQ.~\ref{EQT: hff term simple} gives
\begin{equation}\label{EQT: hff factored}
    2\cproj{\ell_+f_3^\dagg\left(\la_3^\dagg\la_2^- - \widetilde{\la}_3\la_2\right)f_2\ell_+}
\end{equation}
for the APS version and
\begin{equation}\label{EQT: hff factored Pauli}
    \bra{s_3}\left(L_{p_3}^\dagg L_{p_2}^{-1}-(L_{p_3}^{-1})^\dagg L_{p_2}\right)\ket{s_2}
\end{equation}
for the Pauli version. The forms of these equations are identical to the form of the original SA term in EQ.~\ref{EQT: hff term} and imply that the APS scattering amplitude is proportional to
\begin{equation}\label{EQT: hff APS amplitude}
    \mathcal{M}_{hf\overline{f}} \propto \la_3^\dagg\la_2^- - \widetilde{\la}_3\la_2.
\end{equation}
This is the first constructive scattering amplitude to be fully expressed within the APS, coordinate-free, thereby paving the path for all constructive scattering amplitudes---and eventually all of the CSM---to be fully written inside tensor products of the APS. Importantly, the square of this object is 
\begin{equation}\label{EQT: hff Amplitude Squared}
    |\mathcal{M}_{hf\overline{f}}|^2=\mathcal{M}_{hff}\mathcal{M}_{hff}^\dagg\propto 2p_2\sip p_3-2m_f^2,
\end{equation}
where EQ.~\ref{EQW: Spacetime Momentum} was used to simplify the Lorentz spinor products, $p_2\sip p_3$ is the spacetime inner product defined in EQ.~\ref{EQA: SIP}, and $m_f$ is the mass of the fermions. This square, when appropriate multiplication constants are included, is equal to the $hf\overline{f}$ scattering amplitude's square in \cite{Croft2026,Christensen2020}. However, the form of this constructive scattering amplitude within the APS raises questions that must be answered regarding Hermiticity. As a result of the action's Hermiticity in \cite{Christensen2024}, the Hermitian conjugate of a Lagrangian density must look the same after a label-swap.\footnote{It is not a strict equality because the term is then integrated over momentum-space.} However, taking the Hermitian conjugate and then swapping labels instead gives
\begin{equation}\label{EQT: hff Hermiticity}
    \mathcal{M}_{hf\overline{f}}^\dagg |_\text{swap} = \widetilde{\la}_3\la_2 - \la_3^\dagg\la_2^- = -\mathcal{M}_{hff}.
\end{equation}
The $hf\overline{f}$ action term is therefore anti-Hermitian under the CSM's rules. This might look like a problem with the APS, yet looks can be deceiving; it can be proven that the traditional $hf\overline{f}$ action term of \cite{Christensen2024} is also anti-Hermitian. Indeed, this is what SEC.~\ref{D} proves. Further discussion of this anti-Hermiticity will be given in SEC.~\ref{END}. 

\subsection{The $q\overline{q}W$ Term}

Like the $hf\overline{f}$ case, this analysis will take the Lorentz spinors and the massive fields to be respectively form-agnostic so far as they combine to terms of the form $L_{p_j}R_{s_j}$. In fact, all further analyses will do so. Let $u_1$ be the \textit{up}-type quark field, $d_2$ be the \textit{down}-type quark field, and $W_3$ be the W-boson field. Let $\overline{W}_3$ be the oppositely charged W-boson field. From EQ.~$49$ of \cite{Christensen2024}, and once more forgoing the Dirac delta for readability, the $q\overline{q}W$ (W-boson and two quarks) Lagrangian density is
\begin{equation}\label{EQ5F: qqW LD}
    \begin{aligned}
        \mathcal{L}_{q\overline{q}W}&\propto\downSAb{d^\dagg_2\widehat{B}}{\dot{\kk}}{+}\upSAb{\widehat{B}\la_2^\dagg\la_3^-}{\dot{\kk}}{\Le}\downSAb{W_3\widehat{B}}{\Le}{\dot{\mm}}\upSAb{\widehat{B}^\dagg\widetilde{\la}_3\la_1}{\dot{\mm}}{\jj}\downSAb{u_1}{\jj}{+} \\
        &\quad+\downSAb{u^\dagg_2\widehat{B}}{\dot{\kk}}{+}\upSAb{\widehat{B}\la_2^\dagg\la_3^-}{\dot{\kk}}{\Le}\downSAb{\overline{W}_3\widehat{B}}{\Le}{\dot{\mm}}\upSAb{\widehat{B}^\dagg\widetilde{\la}_3\la_1}{\dot{\mm}}{\jj}\downSAb{d_1}{\jj}{+} \\
        &= -2\cproj{\la_3^-W_3\widetilde{\la}_3\la_1u_1\ell_+d_2^\dagg\la^\dagg_2+\la_3^-\overline{W}_3\widetilde{\la}_3\la_1d_1\ell_+u_2^\dagg\la^\dagg_2}.
    \end{aligned}
\end{equation}
It is possible to rearrange the term within the trace such that
\begin{equation}\label{EQ5F: qqW LD 2}
    \mathcal{L}_{q\overline{q}W}\propto -2\cproj{\ell_+d_2^\dagg\la_2^\dagg\la_3^-W_3\widetilde{\la}_3\la_1u_1\ell_++\ell_+u_2^\dagg\la_2^\dagg\la_3^-\overline{W}_3\widetilde{\la}_3\la_1d_1\ell_+},
\end{equation}
which puts the Lagrangian density in the same form as EQ.~\ref{EQT: hff term simple}. But this time, the amplitude cannot be isolated: The Lorentz spinors have an embedded $W_3$. Considering the Hermitian conjugate after relabeling,
\begin{equation}\label{EQ5F: Conjugate & Swap of qqW LD}
    \begin{aligned}
        \mathcal{L}_{q\overline{q}W}^\dagg\big|_{\text{swap}}&\propto -2\cproj{\la_2d_2\ell_+u_1^\dagg\la^\dagg_1\la_3^-W^\dagg_3\widetilde{\la}_3+\la_2u_2\ell_+d_1^\dagg\la^\dagg_1\la_3\overline{W}^\dagg_3\widetilde{\la}_3}\big|_\text{swap} \\
        &= -2\cproj{\la_3^-W_3\widetilde{\la}_3\la_1u_1\ell_+d_2^\dagg\la^\dagg_2+\la_3^-\overline{W}_3\widetilde{\la}_3\la_1d_1\ell_+u_2^\dagg\la^\dagg_2},
    \end{aligned}
\end{equation}
the term is seen to be Hermitian. Notice that the left-chiral Lorentz spinors $\widetilde{\la}_3\la_1$ connect the W-bosons to the quarks and the right-chiral Lorentz spinors $\la_2^\dagg\la_3^-$ connect the W-boson to the antiquarks, thereby representing the left-chiral nature of this weak interaction.

\subsection{The $f\overline{f}Z$ Term}

The $f\overline{f}Z$ (Z-boson and two massive fermions) term is of the same form as the $q\overline{q}W$ (W-boson and two quarks) term, but with an additional \textit{right-chiral} interaction. Let $Z_3$ be the Z-boson field. From EQ.~$49$ of \cite{Christensen2024}, again ignoring the Dirac delta, the Lagrangian density is
\begin{equation}\label{EQ5F: ffZ LD}
    \begin{aligned}
        \mathcal{L}_{f\overline{f}Z}&\propto g_L\downSAb{f^\dagg_2\widehat{B}}{\dot{\kk}}{+}\upSAb{\widehat{B}\la_2^\dagg\la_3^-}{\dot{\kk}}{\Le}\downSAb{Z_3\widehat{B}}{\dot{\mm}}{\Le}\upSAb{\widehat{B}^\dagg\widetilde{\la}_3\la_1}{\dot{\mm}}{\jj}\downSAb{f_1}{\jj}{+} \\
        &\quad+g_R \downSAb{f^\dagg_2\widehat{B}}{\dot{\kk}}{+}\upSAb{\widehat{B}^\dagg\widetilde{\la}_2\la_3}{\dot{\kk}}{\Le}\downSAb{Z_3\widehat{B}}{\dot{\mm}}{\Le}\upSAb{\widehat{B}\la_3^\dagg\la_1^-}{\dot{\mm}}{\jj}\downSAb{f_1}{\jj}{+}\\
        &= -2\cproj{g_L\la_3^-Z_3\widetilde{\la}_3\la_1f_1\ell_+f_2^\dagg\la_2^\dagg+g_R\la_3Z_3\la_3^\dagg\la_1^-f_1\ell_+f_2^\dagg\widetilde{\la}_2},
    \end{aligned}
\end{equation}
where $g_{L/R}\in\BR$ are coupling constants. Being of the same form as EQ.~\ref{EQ5F: qqW LD}, it is Hermitian. In the two separate terms, the chirality of the interaction is encoded by the Lorentz spinors bridging the Z-boson and the fermion field $f_1$.

\subsection{The $l\nu W$ Term}

The $l\overline{\nu}W$ (W-boson, lepton, and antineutrino) is likewise of the same form as both the $q\overline{q}W$ (W-boson and two quarks) and the $f\overline{f}Z$ (Z-boson and two massive fermions) terms. Let $\nu_2$ be the massless neutrino field. From EQ.~$49$ of \cite{Christensen2024}, the Lagrangian density is
\begin{equation}\label{EQ5F: lnW}
    \begin{aligned}
        \mathcal{L}_{l\overline{\nu}W}&\propto\downSAb{\nu^\dagg_{2-}\widehat{B}}{\dot{\kk}}{+}\upSAb{\widehat{B}\la_2^\dagg\la_3^-}{\dot{\kk}}{\Le}\downSAb{W_3\widehat{B}}{\dot{\mm}}{\Le}\upSAb{\widehat{B}^\dagg\widetilde{\la}_3\la_1}{\dot{\mm}}{\jj}\downSAb{f_1}{\jj}{+} \\
        &\quad+\downSAb{f^\dagg_2\widehat{B}}{\dot{\kk}}{+}\upSAb{\widehat{B}\la_2^\dagg\la_3^-}{\dot{\kk}}{\Le}\downSAb{\overline{W}_3\widehat{B}}{\dot{\mm}}{\Le}\upSAb{\widehat{B}^\dagg\widetilde{\la}_3\la_1}{\dot{\mm}}{\jj}\downSAb{\nu_{1-}}{\jj}{+}\\
        &= -2\cproj{\la_3^-W_3\widetilde{\la}_3\la_1f_1\ell_+\nu_{2-}^\dagg\la^\dagg_2+\la_3^-\overline{W}_3\widetilde{\la}_3\la_1\nu_{1-}\ell_+f_2^\dagg\la^\dagg_2}.
    \end{aligned}
\end{equation}
This term is algebraically identical to EQ.~\ref{EQ5F: qqW LD}, but with a massless Lorentz spinors and neutrino fields.

\section{Discussion}\label{END}

The methods of Geometric Algebra have an immensely illuminating power. For many abstract objects within the Constructive Standard Model (CSM) there are now simple geometric interpretations. The constructive amplitudes are simply combinations of Lorentz spinors (standard Lorentz transformations to a particle's momentum) that transform the constructive fields within Lagrangian densities. The spin-$1/2$ fields are simply unnormalized spin rotors ($3$-dimensional rotations that take some reference axis to a particle's spin axis), and spin-$1$ fields are analogous to spin density elements in the limit of their interference terms becoming independent. The Lagrangian densities for $hf\overline{f}$ (Higgs and two massive fermions), $q\overline{q}W$ (W-boson and two quarks), and $l\nu W$ (W-boson, lepton, and neutrino) seen in SEC.~\ref{T} also bear simple geometric interpretations. The former being the measure of the total geometric overlap of $\la_2f_2\ell_+f_3^\dagg\widetilde{\la}_3$ in the timelike direction (spacetime inner product between said term and $1$), and the latter two being effectively interpreted as the spacetime inner products between the spin-$1$ subterms and the combined spin-$1/2$ subterms. Moreover, the constructive fields are either equivalent to or directly related to objects within Pauli Theory and Quantum Information \cite{McKenzie2015,Burns2025,Lounesto2001,Doran2003,Sobczyk2019,Vaz2019}---so advances in one field are immediately relevant to the other. 

 Yet, in spite of these advances, there are still kinks to be worked out. While the fields of SEC.~\ref{F} exist without issue inside the Algebra of Physical Space (APS), the massless spin-$1$ fields interact with an entity known as the $x$-factor \cite{Christensen2023}---whose form within the SA has yet to be explored. This is currently being investigated and will be the subject of a future paper. Once this wrinkle is ironed out, it is expected that many significant simplifications will follow. This is generally motivated by the conceptual and mathematical simplicity that the APS has already afforded, but more specifically motivated by the discovery of the first constructive amplitude (EQ.~\ref{EQT: hff APS amplitude}) to be \textit{fully inside the APS}. 

 With mention of the first APS amplitude, it is important to discuss the anti-Hermiticity of its associated action.  In the CSM, an action contribution is called Hermitian if the Hermitian conjugate of the Lagrangian density within the integral, after appropriate relabeling of particle momenta, looks the same as the original term. This definition is used due to the Lagrangian densities existing within integrals over momentum space. When this is performed on the APS amplitude for $hf\overline{f}$ (Higgs and two massive fermions), as seen in EQ.~\ref{EQT: hff Hermiticity}, an overall minus sign appears. Therefore the amplitude---and so the action term---is anti-Hermitian. One might argue that this is a defect of using the APS for the CSM, however SEC.~\ref{D} proves the anti-Hermiticity using traditional CSM techniques. So, the anti-Hermiticity appears to be a true feature of the CSM. The interpretation of this is an open question, and must be resolved. 

In addition to everything previously mentioned, this work also explicitly established \textit{Wigner-covariance} in the APS. These derivations, when compared to traditional derivations \cite{Christensen2024,Parmar2021,DeLaurentis2026}, are notably shorter and in the achiral case are fully coordinate-free. When all Lorentz spinors and their transformations were injected into the SA, exact equivalents of equations to those in the traditional CSM were obtained. This is further motivation that in time everything from the CSM might be treated in a wholly geometric and coordinate-free way.

In short, this work---from the Wigner-covariance to the construction of fields and preliminary Lagrangian densities---extended the methods and results of \cite{Croft2026}, further establishing the usefulness of Geometric Algebra-based Physics. However, work regarding the massless spin-$1$ case remains. This is currently being examined and will be the subject of future work.


\section*{Acknowledgments}
The authors would like to thank Edward Corbett, Martin Roelfs, David Eelbode, and Eren Erdo\u{g}an for their support and helpful discussions. The authors would also like to express their gratitude to the American people for partially supporting this work through the National Science Foundation under Grant No. PHY-2411482.

\subsubsection*{Data Availability Statement}
This paper contains no results found through data analysis or any like method.


\appendix

\newpage

\section{Essential Conventions}\label{A}

This paper makes use of the \textit{Algebra of Physical Space} (APS), which is the real geometric algebra $\APS$. Let $g,h\in\APS$ be arbitrary multivectors in the APS. Within the notational convention of this paper, \textit{grade involution} is
\begin{equation}\label{EQA: Parity Conjugation}
    g^- = \gproj{g}{0}-\gproj{g}{1}+\gproj{g}{2}-\gproj{g}{3},
\end{equation}
and is called \textit{parity conjugation}. This is because grade involution negates all basis directions of $3$-dimensional space. Then \textit{reversion} is
\begin{equation}\label{EQA: Hermitian Conjugation}
    g^\dagg = \gproj{g}{0}+\gproj{g}{1}-\gproj{g}{2}-\gproj{g}{3},
\end{equation}
and $(gh)^\dagg=h^\dagg g^\dagg$. This is called \textit{Hermitian conjugation} as that is to what reversion within $3$-dimensional space corresponds. Lastly, the combination of parity and Hermitian conjugation is \textit{Clifford conjugation}:
\begin{equation}\label{EQA: Clifford Conjugation}
    \widetilde{g} = (g^-)^\dagg = (g^\dagg)^- = \gproj{g}{0}-\gproj{g}{1}+\gproj{g}{2}-\gproj{g}{3}.
\end{equation}
If $A\in\APS^j$ is grade-$j$, then
\begin{equation}\label{EQA: Magnitude}
    |A|=\sqrt{AA^\dagg}
\end{equation}
is the definition of its \textit{magnitude}.

\subsection{Spacetime in $\APS$}

The APS may describe $(1+3)$-spacetime geometry because it is isomorphic to the even subalgebra of the \textit{Spacetime Algebra}, $\APS\approx\STA^+$. The sum of scalars and vectors within the APS is called a \textit{paravector} and fills the role of a spacetime vector:
\begin{equation}\label{EQA: Paravector}
    a = \gproj{a}{0}+\gproj{a}{1} = a_t +\mathbf{a}.
\end{equation}
For a Lorentz transformation $\Lambda\in\Spin(1,3)\subset\APS$, a paravector is Lorentz-covariant via
\begin{equation}\label{EQA: Lorentz Transformation}
    a\mapsto \Lambda a\Lambda^\dagg.
\end{equation}
The product of a paravector with its Clifford conjugate is a Lorentz-invariant scalar,
\begin{equation}\label{EQA: Paravector squared}
    a\widetilde{a}=\widetilde{a}a=aa^-=a^-a=a_t^2-\mathbf{a}^2.
\end{equation}
Given another paravector $b\in\APS^{0\oplus3}$, the \textit{spacetime product} is
\begin{equation}\label{EQA: Spacetime Product}
    a\widetilde{b} = a\sip b+a\sop b.
\end{equation}
The first product is the \textit{spacetime inner product},
\begin{equation}\label{EQA: SIP}
    a\sip b = \frac{1}{2}\left(a\widetilde{b}+b\widetilde{a}\right) = a_tb_t-\mathbf{a}\cdot\mathbf{b},
\end{equation}
and the second product is the \textit{spacetime outer product},
\begin{equation}\label{EQA: SOP}
    a\sop b = \frac{1}{2}\left(a\widetilde{b}-b\widetilde{a}\right)= b_t\mathbf{a}-a_t\mathbf{b}-\mathbf{a}\wedge\mathbf{b}.
\end{equation}
There is also an opposite parity definition for the spacetime product and its sub-products, obtained through EQ.~\ref{EQA: Parity Conjugation}. 

\section{Spinor Formalism Recapitulation}\label{B}

The previous paper introduced a wholistic spinor formalism \cite{Croft2026}. This formalism dealt with Pauli spinors, Lorentz spinors, and ray spinors. Each of these will now be briefly summarized.

\subsection{Pauli Spinors}

Let $\hat{a}, \hat{s},\hat{p}\in\APS^1$ respectively be some unit \textit{reference axis}, some unit \textit{spin vector}, and some unit \textit{spatial-momentum direction}. Any rotor $R_s\in\Spin(3)\subset\APS$ such that
\begin{equation}\label{EQB: R_s}
    \hat{s}=R_s\hat{a}R_s^\dagg
\end{equation}
is called the \textit{spin rotor with respect to} $\hat{a}$. This rotor is also called the \textit{operatorial Pauli spinor}. If
\begin{equation}\label{EQB: spin +-1}
    \hat{s}\cdot\hat{a}=\pm1,
\end{equation}
then, without loss of generality, $R_s=1$ is called \textit{spin-up with respect to} $\hat{a}$ and $R_s=\hat{B}\in\APS^2$ is called \textit{spin-down with respect to} $\hat{a}$. Said otherly: If $R_s=1$, then $\hat{s}\cdot\hat{a}=+1$. And if $R_s=\hat{B}$, then $\hat{s}\cdot\hat{a}=-1$.

\subsection{Ray Spinor Structure}

Let $\hat{a}\in\APS^1$ be an arbitrary unit \textit{reference axis}. Then the idempotents (self-squaring elements)
\begin{equation}\label{EQB: Idempotents}
    \ell_\pm = \frac{1}{2}(1\pm\hat{a})
\end{equation}
geometrically describe lightlike trajectories, earning them the title of \textit{lightrays} in the direction of $\pm\hat{a}$. This is because the unit scalar represents the unit \textit{timelike} direction and the unit reference axis represents a unit \textit{spacelike} direction. Using the Clifford conjugate, which induces the Minkowski metric inside the APS,
\begin{equation}\label{EQB: Nilpotent}
    \widetilde{\ell}_\pm\ell_\pm=\ell_\mp\ell_\pm = 0,
\end{equation}
the lightrays are indeed lightlike (zero-squaring under the Minkowski metric). Furthermore,
\begin{equation}\label{EQB: Partition Unity}
    1=\ell_++\ell_-,
\end{equation}
the lightrays \textit{partition unity} and are called \textit{complementary}. The above is an algebraic statement that the lightrays border the lightcone and partition the timelike direction. By these lightrays, any multivector $g\in\APS$ can be \textit{lightray-partitioned}, 
\begin{equation}\label{EQB: LR-Partition}
    g = g\ell_++g\ell_- = \al_\Fo+\be_\Ba.
\end{equation}
Respectively, $\al_\Fo=g\ell_+$ and $\be_\Ba=g\ell_-$ are called \textit{forward-oriented} and \textit{backward-oriented} (light)\textit{ray} spinors. For simplicity, these are each shortened to F-ray spinor and B-ray spinor. \textit{Minimal left/right ideals} are minimal subsets of an algebra wherein an element of the ideal, when multiplied by an element of the algebra on the left/right, stays within the ideal. F-ray spinors are elements of the minimal left ideal $\APS\ell_+$ while B-ray spinors are elements of the complementary $\APS\ell_-$. The minimal right ideals $\ell_+\APS$ and $\ell_-\APS$ are also used within ray spinor structure. Let $\hat{b}\in\APS^1$ be a unit vector such that $\hat{a}\cdot\hat{b}=0$ and $\hat{B}\ell_\pm=\pm\hat{b}\ell_\pm$. Then \textit{dual} ray spinors are defined as
\begin{equation}\label{EQB: Dual F-ray}
    \al^\Fo = \hat{b}\widetilde{\al}_\Fo=\ell_+\hat{b}\widetilde{g}\in\ell_+\APS
\end{equation}
and
\begin{equation}\label{EQB: Dual B-ray}
    \be^\Ba = \hat{b}\widetilde{\be}_\Ba=\ell_-\hat{b}\widetilde{g}\in\ell_-\APS.
\end{equation}
Products between dual ray spinors and ray spinors are Lorentz-invariant. Supposing that $\psi_{\Fo/\Ba}$ and $\phi_{\Fo/\Ba}$ are arbitrary ray spinors of identical orientation, then together they form the \textit{symplectic} Lorentz-invariant product
\begin{equation}\label{EQB: Symplectic}
    \psi^{\Fo/\Ba}\phi_{\Fo/\Ba}=\hat{b}\widetilde{\psi}_{\Fo/\Ba}\phi_{\Fo/\Ba}.
\end{equation}
Supposing that $\psi_{\Ba/\Fo}$ and $\phi_{\Fo/\Ba}$ are arbitrary ray spinors of opposing orientation, then they instead form 
\begin{equation}\label{EQB: Non-symplectic}
    \psi^{\Ba/\Fo}\phi_{\Fo/\Ba}=\hat{b}\widetilde{\psi}_{\Ba/\Fo}\phi_{\Fo/\Ba},
\end{equation}
the \textit{non-symplectic} Lorentz-invariant product.

\subsection{The Scattering Algebra}

The \textit{Scattering Algebra} (SA) is an algebraic tool \textit{inside} the Algebra of Physical Space (APS) that connects to the helicity methods of the Constructive Standard Model. If $g\in\APS$ is an arbitrary multivector, then it may be injected into its \textit{SA bracket}:
\begin{equation}\label{EQB: SA Bracket}
    g\longmapsto\upSAb{g}{\jj}{\kk}=\ell_{\jj} g\ell_{\kk}+\hat{b}\ell_{\jj} g\ell_{\kk}\hat{b}.
\end{equation}
The indices $\jj,\kk\in\{+,-\}$ are over lightray-orientation, and $\hat{b}\in\APS^1$ is the appropriate unit vector defined in SEC.~\ref{B} satisfying $\hat{a}\cdot\hat{b}=0$. The Hermitian conjugate of the SA bracket is 
\begin{equation}\label{EQB: SA Bracket Conj.}
    \left(\upSAb{g}{\jj}{\kk}\right)^\dagg=\downSAb{g^\dagg}{\jj}{\kk}=\ell_{\kk} g^\dagg\ell_{\jj}+\hat{b}\ell_{\kk} g^\dagg\ell_{\jj}\hat{b}.
\end{equation}
The convention of raising and lowering indices for such SA brackets is solely visual, allowing for the Einstein summation convention. Using this convention, and for multivectors $g_j\in\APS$,
\begin{equation}\label{EQB: Central Proj. Collapse}
    \upSAb{g_1}{\mathrm{A}}{\mathrm{B}}\downSAb{g_2}{\mathrm{C}}{\mathrm{B}}\upSAb{g_3}{\mathrm{C}}{\mathrm{D}}\cdots=2\cproj{g_1g_2g_3\dots}.
\end{equation}
Here $\cproj{\cdot}$ is the \textit{central projection}\footnote{Projection to the grade-$0$ and grade-$3$ subspaces of the APS.} and the final index of summation is a lowered $\mathrm{A}$. This is equivalent to a traditional matrix-trace.

To emphasize that the indices' positions are but convention, consider the following "redefinition" of the SA bracket:
\begin{equation}\label{EQB: Redef. of SA Bracket}
    \SAb{g}_{\al}^{\ \jj} =\ell_{\al} g\ell_{\jj}+\hat{b}\ell_{\al} g\ell_{\jj}\hat{b}.
\end{equation}
Its Hermitian conjugate then becomes
\begin{equation}\label{EQB: Redef. of SA Bracket Conj.}
    \left(\SAb{g}_{\al}^{\ \jj}\right)^\dagg = \SAb{g^\dagg}_{\jj}^{\ \al} =\ell_{\jj} g^\dagg\ell_{\al}+\hat{b}\ell_{\jj} g^\dagg\ell_{\al}\hat{b}.
\end{equation}
This results in EQ.~\ref{EQB: Central Proj. Collapse} taking the form
\begin{equation}\label{EQB: Redef. Collapse}
    \SAb{g_1}_{\al}^{\ \jj}\SAb{g_2}_{\jj}^{\ \be}\SAb{g_3}_{\be}^{\ \kk}\cdots=2\cproj{g_1g_2g_3\dots},
\end{equation}
which is functionally identical. Such different definitions of the SA bracket are called \textit{lightray filters}, or simply \textit{filters}. Geometrically, each definition filters a multivector's components via lightray-alignment. One might think that swapping between index convention is impractical, but it allows for a \textit{direct} rigorous connection between Lorentz spinors of the APS and angle spinors\footnote{And a \textit{transpose} rigorous connection between Lorentz spinors and square spinors. This will be explored more thoroughly in an upcoming paper.} of the Constructive Standard Model. Indeed, SEC.~\ref{W} makes use of EQ.~\ref{EQB: Redef. of SA Bracket}.

\section{Proof of $hf\overline{f}$ Anti-Hermiticity}\label{D}

Following \cite{Christensen2024}, in the Constructive Standard Model (CSM) a Lagrangian density's contribution to the action is considered \textit{Hermitian} if its Hermitian conjugate looks the same as the original term after a label-swap. The reason for this peculiar definition is that these terms appear in integrals over momentum-space and the indices are dummy integration indices. It was shown in SEC.~\ref{T} that the Algebra of Physical Space's version of the constructive $hf\overline{f}$ amplitude---and therefore the $hf\overline{f}$ action term---is anti-Hermitian. This can also be shown using the traditional methods of the CSM.

Ignoring the Higgs field $h(p_1)$, the traditional Lagrangian density term is linearly proportional to
\begin{equation}\label{EQD: CSM Action Term}
    \mathcal{L}_{hff} \propto \overline{f}_{2\, \kk}\left(\langle\mathbf{32}\rangle^{\jj\kk}+\left[\mathbf{32}\right]^{\jj\kk}\right)f_{3\, \jj},
\end{equation}
where $2$ and $3$ are labels for the fermion with momentum $p_2$ and its own antifermion with momentum $p_3$. The fields are defined as the ordered sets
\begin{equation}\label{EQD: Field Sets}
    f_{3}^{\jj}\equiv(f_3^\uparrow,f_3^\downarrow)\quad\text{and}\quad \overline{f}_{2\,\jj}\equiv(\overline{f}_2^\uparrow,\overline{f}_2^\downarrow)
\end{equation}
such that they satisfy the Pauli inner product 
\begin{equation}\label{EQD: Pauli Product}
    f_3^{\jj}\overline{f}_{2\,\jj} = f_3^\uparrow\overline{f}_2^\uparrow+f_3^\downarrow \overline{f}_2^\downarrow.
\end{equation}
This implies that for the epsilon tensors,
\begin{equation}\label{EQD: Epsilon Tensor}
    \epsilon_{\jj\kk}\equiv\begin{bmatrix}
        0 & -1 \\
        1 & 0
    \end{bmatrix}\quad\text{and}\quad\epsilon^{\jj\kk}\equiv\begin{bmatrix}
        0 & 1 \\
        -1 & 0
    \end{bmatrix},
\end{equation}
the fields with lowered/raised indices are
\begin{equation}\label{EQD: Field Index Relations}
    f_{3\,\jj} = f_{3}^\mathrm{A}\epsilon_{\mathrm{A}\jj}\equiv(f_3^\downarrow,-f_3^\uparrow)\quad\text{and}\quad \overline{f}_{2}^{\jj}=\overline{f}_{2\,\mathrm{A}}\epsilon^{\mathrm{A}\jj}\equiv(-\overline{f}_2^\downarrow,\overline{f}_2^\uparrow).
\end{equation}
And the Hermitian conjugates of the lower-index fields are
\begin{equation}\label{EQD: Field Sets HC}
    (f_{3}^\dagger)^{\jj}\equiv\left((f_3^\downarrow)^\dagg,-(f_3^\uparrow)^\dagg\right)\quad\text{and}\quad (\overline{f}_{2}^\dagg)^{\jj}\equiv\left((\overline{f}_2^\uparrow)^\dagg,(\overline{f}_2^\downarrow)^\dagg\right).
\end{equation}
Taking the Hermitian conjugate of EQ.~\ref{EQD: CSM Action Term} gives
\begin{equation}\label{EQD: CSM Action HC}
    \begin{aligned}
        \mathcal{L}_{hff}^\dagg&\propto (f_3^\dagg)^{\jj}\left(\left[\mathbf{23}\right]_{\kk\jj}+\langle\mathbf{23}\rangle_{\kk\jj}\right)(\overline{f}_2^\dagg)^{\kk} \\
        &=(f_3^\dagg)^{\jj}\delta_{\jj}^{\nn}\left(\left[\mathbf{23}\right]_{\mathrm{P}\nn}+\langle\mathbf{23}\rangle_{\mathrm{P}\nn}\right)\delta_{\kk}^\mathrm{P}(\overline{f}_2^\dagg)^{\kk} \\
        &=(f_3^\dagg)^{\jj}\epsilon_{\jj\mm}\epsilon^{\mm\nn}\left(\left[\mathbf{23}\right]_{\mathrm{P}\nn}+\langle\mathbf{23}\rangle_{\mathrm{P}\nn}\right)\epsilon_{\kk\Ri}\epsilon^{\Ri\mathrm{P}}(\overline{f}_2^\dagg)^{\kk} \\
        &=\left(f_{3}^\dagg\right)_{\mm}\left(\langle\mathbf{23}\rangle^{\Ri\mm}+\left[\mathbf{23}\right]^{\Ri\mm}\right)\left(\overline{f}_{2}^\dagg\right)_{\Ri}.
    \end{aligned}
\end{equation}
Naïvely, the action term seems Hermitian under the relabelings $\mathbf{2}\mapsto\mathbf{3}$, $\mathbf{3}\mapsto\mathbf{2}$, $f_3^\dagg\mapsto \overline{f}_2$, and $\overline{f}_2^\dagg\mapsto f_3$. But expanding the fields into their transformed ordered sets reveals an overall minus sign:
\begin{equation}\label{EQD: Field Set Transforms}
    \begin{aligned}
        &(f_3^\downarrow,-f_3^\uparrow)\xrightarrow{(\cdot)^\dagg}\left((f_3^\uparrow)^\dagg,-(f_3^\downarrow)^\dagg\right)\xrightarrow{(\cdot)\epsilon_{\jj\mm}}(-(f_3^\uparrow)^\dagg,-(f_3^\downarrow)^\dagg) \\
        &(\overline{f}_2^\uparrow,\overline{f}_2^\downarrow)\xrightarrow{(\cdot)^\dagg}\left((\overline{f}_2^\uparrow)^\dagg,(\overline{f}_2^\downarrow)^\dagg\right)\xrightarrow{(\cdot)\epsilon_{\kk\Ri}}\left((\overline{f}_2^\downarrow)^\dagg,-(\overline{f}_2^\uparrow)^\dagg\right).
    \end{aligned}
\end{equation}
That is, the Lagrangian density will only return to "looking the same" under the relabelings $\mathbf{2}\mapsto\mathbf{3}$, $\mathbf{3}\mapsto\mathbf{2}$, $f_3^\dagg\mapsto -\overline{f}_2$, and $\overline{f}_2^\dagg\mapsto f_3$. With an overall minus sign, this means the action term must be anti-Hermitian. It will not be shown here, but it is also possible to prove this result within the Scattering Algebra, before the collapse to the central projection.

\vspace*{2mm}

\section{Table of Concurrence}\label{E}
\noindent
\begin{center}
\begin{tabular}{llll}
\toprule
\textbf{Lorentz Spinors} & \textbf{Traditional} &  \textbf{SA} & \textbf{APS} \\
\midrule
Right-Angle Spin Spinor & $|\mathbf{j}\rangle_\al^{\ \jj}$ & $\SAb{\la}_\al^{\ \jj}$ & $\la=\sqrt{mc}R_pL_a$\\
Left-Angle Spin Spinor* & $\langle\mathbf{j}|^{\al\jj}$ & $\SAb{\hat{B}\widetilde{\la}}^{\al\jj}$ & $\hat{B}\widetilde{\la}$\\
Right-Square Spin Spinor & $|\mathbf{j}\rbrack^{\dot{\be}}_{\ \jj}$ & $\SAb{\la^-\hat{B}^\dagg}^{\dot{\be}}_{\ \jj}$ & $\la^-\hat{B}^\dagg$\\
Left-Square Spin Spinor* & $\lbrack\mathbf{j}|_{\dot{\be}\jj}$ & $\SAb{\la^\dagg}_{\dot{\be}\jj}$ & $\la^\dagg$\\
(Raising) Epsilon Tensor & $\epsilon^{\jj\kk}$ & $\upSAb{\hat{B}^\dagg}{\jj}{\kk}$ & $\hat{B}^\dagg$\\
(Lowering) Epsilon Tensor & $\epsilon_{\jj\kk}$ & $\downSAb{\hat{B}}{\kk}{\jj}$ & $\hat{B}$\\
(Left) Lorentz Product & $\langle\mathbf{j}|^{\al\jj}|\mathbf{k}\rangle_\al^{\ \kk}$ & $\upSAb{\LP{j}{k}}{\jj}{\kk}$ & $\LP{j}{k}$ \\
(Right) Lorentz Product & $\lbrack\mathbf{j}|_{\dot{\be}}^{\ \jj}|\mathbf{k}\rbrack^{\dot{\be}\kk}$ & $\upSAb{\mLP{j}{k}}{\jj}{\kk}$ & $\mLP{j}{k}$\\
Right-Angle Helicity Spinor & $|j\rangle_\al$ & $\SAb{\la_\Fo}_\al^{\ +}$ & $\la_\Fo=\sqrt{\frac{2E}{c}}R_p\ell_+$\\
Left-Angle Helicity Spinor* & $\langle j|^{\al}$ & $\SAb{-\la^\Fo}^{\al+}$ & $\la^\Fo=\hat{b}\widetilde{\la}_\Fo$\\
Right-Square Helicity Spinor & $| j\rbrack^{\dot{\be}}$ & $\SAb{-(\la^\Fo)^\dagg}^{\dot{\be}}_{\ +}$ & $(\la^\Fo)^\dagg$\\
Left-Square Helicity Spinor* & $\lbrack j|_{\dot{\be}}$ & $\SAb{\la_\Fo^\dagg}_{\dot{\be}+}$ & $\la_\Fo^\dagg$\\
\bottomrule
\end{tabular}
\end{center}
All above labels with an attached asterisk (*) denote objects whose traditional representation and representation with the Scattering Algebra (SA) is the \textit{transpose} of the matrix representation for the object in the Algebra of Physical Space (APS).
\noindent
\begin{center}
\begin{tabular}{llll}
\toprule
\textbf{Constructive Fields} & \textbf{Traditional} &  \textbf{SA} & \textbf{APS} \\
\midrule
Lowered Massive Spin-$1/2$ & $f_\jj$ & $\downSAb{f'}{+}{\jj}$ & $f'=f_\uparrow R_p^\dagg-f_\downarrow R_p^\dagg\hat{B}$\\
Lowered Massive Anti-Spin-$1/2$ & $\overline{f}_\jj$ & $\SAb{f'^\dagg\hat{B}}_{+\jj}$ & $f'^\dagg\hat{B}$\\
Lowered Massive Spin-$1$ & $A_{\jj\kk}$ & Unknown & Defined in EQ.~\ref{EQF: Primed massive spin-one field}\\
Spin-$1/2$ Helicity-$\uparrow$** & $\nu_\uparrow$ & $\SAb{\nu_\uparrow}_{++}$ & $\nu_\uparrow$\\
Spin-$1$ Helicity-$\uparrow$** & $A_\uparrow$ & Unknown & $A_\uparrow\ell_+$ \\
\bottomrule
\end{tabular}
\end{center}
All above labels with an attached double-asterisk (**) denote fields that are only being compared for a specific helicity orientation.
\noindent
\begin{center}
\begin{tabular}{llll}
\toprule
\textbf{Wigner Little Group Rotations} & \textbf{Traditional} &  \textbf{SA} & \textbf{APS} \\
\midrule
Massive Rotation & $\left(e^{i\harpoon \omega\cdot\harpoon J}\right)_\kk^{\ \jj}$ & $\SAb{S}_{\kk}^{\ \jj}$ & $S=L_{\overline{a}}^-R_{\overline{p}}^\dagg\Lambda R_pL_a$\\
Massless Rotation & $e^{-i\frac{1}{2}\omega}$ & $e^{-i\frac{1}{2}\omega}$ & $e^{-i\frac{1}{2}\omega\hat{a}}=R_{\overline{p}}^\dagg RR_p$\\
\bottomrule
\end{tabular}
\end{center}
\printbibliography


\end{document}